\documentclass[apj,twocolumn]{openjournal}
\usepackage{natbib} 
\usepackage[dvipsnames]{xcolor}
\usepackage{aas_macros} 
\usepackage{amssymb}
\usepackage{amsmath}
\usepackage[title]{appendix}
\usepackage{hyperref}	
\hypersetup{colorlinks=true,linkcolor=blue,citecolor=blue,filecolor=blue,urlcolor=blue}
\usepackage[caption=false]{subfig}
\usepackage{soul}                          
\usepackage{xspace}
\usepackage{booktabs}
\usepackage{savesym}
\savesymbol{tablenum}
\usepackage{siunitx}
\usepackage{pifont}
\usepackage{float}
\usepackage{lipsum}
\usepackage{bm}
\restoresymbol{SIX}{tablenum}

\makeatletter 
  \patchcmd{\NAT@citex}
    {\@citea\NAT@hyper@{%
      \NAT@nmfmt{\NAT@nm}%
      \hyper@natlinkbreak{\NAT@aysep\NAT@spacechar}{\@citeb\@extra@b@citeb}%
      \NAT@date}}
    {\@citea\NAT@nmfmt{\NAT@nm}%
    \NAT@aysep\NAT@spacechar\NAT@hyper@{\NAT@date}}{}{}

  \patchcmd{\NAT@citex}
    {\@citea\NAT@hyper@{%
      \NAT@nmfmt{\NAT@nm}%
      \hyper@natlinkbreak{\NAT@spacechar\NAT@@open\if*#1*\else#1\NAT@spacechar\fi}%
        {\@citeb\@extra@b@citeb}%
      \NAT@date}}
    {\@citea\NAT@nmfmt{\NAT@nm}%
    \NAT@spacechar\NAT@@open\if*#1*\else#1\NAT@spacechar\fi\NAT@hyper@{\NAT@date}}
    {}{}
\makeatother

\DeclareSIUnit\Msun{\ensuremath{M_\odot}}
\DeclareSIUnit\Zsun{\ensuremath{Z_\odot}}
\DeclareSIUnit\hred{\ensuremath{\textit{h}}}

\definecolor{OG}{RGB}{36, 104, 126}
\definecolor{SF}{RGB}{154, 187, 107}
\definecolor{IMF}{RGB}{242, 148, 13}
\definecolor{ISM}{RGB}{167, 28, 71}

\newcommand{\OG}[0]{\textit{efficient star formation}}
\newcommand{\SF}[0]{\textit{bursty star formation}}
\newcommand{\IMF}[0]{\textit{variable IMF}}
\newcommand{\ISM}[0]{\textit{high $\epsilon_{\rm ff}$, HN}}

\begin{document}

\title[]{
Anatomy of LyC Escape in MEGATRON: Connecting Feedback Physics to Demographics, Leakage Timescales, and Explorations of ISM Density as a Tracer of Escape Fraction
\vspace{-15mm}}
\author{Nicholas Choustikov$^{1,2,3*}$,
Harley Katz$^{4,5}$,
Taysun Kimm$^{6}$,
Matthew J. Hayes$^{7}$,
Aaron Smith$^{8}$, \\
Crystal L. Martin$^{9}$,
Shyam H. Menon$^{10,11}$,
Larrance Xing$^{4}$,
Joki Rosdahl$^{12}$,
John Chisholm$^{13,14}$,\\
Aayush Saxena$^{3,15}$,
Martin P. Rey$^{16}$,
Adrianne Slyz$^{3}$,
Julien Devriendt$^{3}$,
Alex J. Cameron$^{17,18}$,\\
Anatole Storck$^{3}$,
and Corentin Cadiou$^{19}$}
\thanks{$^*$E-mail: \href{nc538@cam.ac.uk}{nc538@cam.ac.uk}}
\affiliation{$^{1}$Kavli Institute for Cosmology, University of Cambridge, Madingley Road, Cambridge CB3 0HA, UK}
\affiliation{$^{2}$Institute of Astronomy, University of Cambridge, Madingley Road, Cambridge CB3 0HA, UK}
\affiliation{$^{3}$Sub-department of Astrophysics, University of Oxford, Keble Road, Oxford OX1 3RH, United Kingdom}
\affiliation{$^{4}$Department of Astronomy \& Astrophysics, University of Chicago, 5640 S Ellis Avenue, Chicago, IL 60637, USA}
\affiliation{$^{5}$Kavli Institute for Cosmological Physics, University of Chicago, Chicago IL 60637, USA}
\affiliation{$^{6}$Department of Astronomy, Yonsei University, 50 Yonsei-ro, Seodaemun-gu, Seoul 03722, Republic of Korea}
\affiliation{$^{7}$Stockholm University, Department of Astronomy and Oskar Klein Centre for Cosmoparticle Physics, AlbaNova University Centre, SE-10691, Stockholm, Sweden}
\affiliation{$^{8}$Department of Physics, The University of Texas at Dallas, Richardson, TX 75080, USA}
\affiliation{$^{9}$Department of Physics, University of California Santa Barbara, Santa Barbara, CA 93106, USA}
\affiliation{$^{10}$Centre for Computational Astrophysics, Flatiron Institute, 162 5th Avenue, New York, NY 10010, USA}
\affiliation{$^{11}$Department of Physics and Astronomy, Rutgers University, 136 Frelinghuysen Road, Piscataway, NJ 08854, USA}
\affiliation{$^{12}$Universite Claude Bernard Lyon 1, CRAL UMR5574, ENS de Lyon, CNRS, Villeurbanne, F-69622, France}
\affiliation{$^{13}$Department of Astronomy, The University of Texas at Austin, Austin, TX 78712, USA}
\affiliation{$^{14}$Cosmic Frontier Center, The University of Texas at Austin, Austin, TX 78712, USA}
\affiliation{$^{15}$Department of Physics and Astronomy, University College London, Gower Street, London WC1E 6BT, United Kingdom}
\affiliation{$^{16}$University of Bath, Department of Physics, Claverton Down, Bath, BA2 7AY, UK}
\affiliation{$^{17}$Cosmic Dawn centre (DAWN), Copenhagen, Denmark}
\affiliation{$^{18}$Niels Bohr Institute, University of Copenhagen, Jagtvej 128, DK-2200 Copenhagen, Denmark}
\affiliation{$^{19}$Institut d'Astrophysique de Paris, Sorbonne Université, CNRS, UMR 7095, 98 bis bd Arago, 75014 Paris, France}

\begin{abstract}
\noindent
Constraining the Lyman Continuum (LyC) escape fraction ($f_{\rm esc}$) of high-redshift galaxies is essential to connect the abundant populations revealed by \textit{JWST} to cosmic reionization. We investigate this interplay using {\small MEGATRON}, a suite of zoom simulations with detailed radiation/chemistry and varied star-formation/feedback prescriptions. We find that feedback physics simultaneously regulates UV source counts, LyC escape fractions, and escaping ionizing luminosities ($\dot{N}_{\rm ion,esc}$), producing markedly different reionization histories. These differences are governed primarily by the delay between a starburst and LyC escape: efficient early feedback and lower-density star-forming environments enable photons to escape while pre-supernovae stellar populations remain strongly ionizing. Consequently, models that produce fewer UV-bright galaxies yield a shorter delay for LyC escape and reionize earlier than models with more abundant UV sources and delayed LyC emission. Furthermore, UV-faint galaxies follow a common physical sequence in ISM structure with increasing $f_{\rm esc}$ irrespective of feedback physics: stronger leakers ($1\%\leq f_{\rm esc}\leq 50\%$) exhibit evidence of strong photoionization and SNe feedback alongside post-starburst star formation histories (SFHs), while extreme leakers ($f_{\rm esc}\gtrsim50\%$) have also cleared their cold/dense ISM. We find that electron densities inferred from the $\rm [O\,{II}]\,\lambda\lambda3727$ doublet are a candidate diagnostic of $f_{\rm esc}$ in UV-faint galaxies, with lower densities around $\rm H~{\small II}$ regions selecting stronger leakers. Moreover, this density can trace the early evolution of $f_{\rm esc}$, whereas UV slope-based calibrations over-predict $\dot{N}_{\rm ion,esc}$ for individual young galaxies by up to two orders of magnitude. Overall, the delayed duty cycle of LyC leakage must be accounted for when UV source counts are interpreted in the context of reionization. Resolved ISM diagnostics and detailed SFHs offer a promising route to do so.

\end{abstract}

\section{Introduction}

\noindent The Epoch of Reionization describes the period during which neutral hydrogen in the intergalactic medium (IGM) became ionized by the first galaxies. This put an end to the dark ages which had long dominated the Universe for the first billion years ($z\sim 5.3$, \citealt{Robertson2015,Kulkarni2019,Keating:2020,Becker:2021}). A variety of probes have been used to constrain this process, including the existence of Gunn-Peterson troughs in quasar spectra \citep{Keating:2020,Becker:2021,Bosman2022}, the damping wings of high-redshift quasars \citep[][]{Davies:2018,Durovcikova:2020}, the decrease in number density of $\rm Ly\alpha$ emitters (LAEs, \citealt{Stark:2010,Pentericci:2011,Mason:2018,Mason:2019}), the opacity of the $\rm Ly\alpha$ forest of quasars \citep{Fan:2006,Fan:2006b} and galaxies \citep{Meyer:2025}, and measurements from the CMB \citep[e.g.,][]{Heinrich:2021}.

The sources of ionizing Lyman Continuum (LyC) photons remain largely debated, with various models appealing to differing contributions from UV-bright sources \citep[e.g.,][]{Naidu2020} and abundant UV-faint galaxies \citep[e.g.,][]{Finkelstein2019}. In either case, typical models for reionization driven by star-forming galaxies rely on three quantities, which together provide an estimate of the total ionizing photon production rate, given by $\dot{n}_{\rm ion} = \rho_{\rm UV}\xi_{\rm ion}f_{\rm esc}$ (see review by \citealt{Robertson:2022}). These include the redshift-dependent total UV luminosity density ($\rho_{\rm UV}$), the ionizing photon production efficiency ($\xi_{\rm ion}$), giving the number of ionizing photons produced by a stellar population per UV luminosity, and the LyC escape fraction ($f_{\rm esc}$), reporting the number of these photons which escape their host galaxy. Another approach is to combine these quantities and to instead try to constrain the `escaping ionizing luminosity' ($\dot{N}_{\rm ion,esc}$) of individual galaxies (see discussion by \citealt{Choustkov:2025}). Once estimated for each source and aggregated across a survey volume, the LyC production rate can be estimated by computing an ionizing emissivity function, $\phi_{\rm ion}(\dot{N}_{\rm ion,esc}, z)$, which can then be integrated to compute $\dot{n}_{\rm ion}$ \citep[e.g.,][]{Giovinazzo:2026b,Summerfield:2026}.

Since the launch of the \textit{James Webb Space Telescope} (\textit{JWST}), much effort has gone to placing significant constraints on each of these quantities. In particular, many studies have tried to constrain the high-redshift UV luminosity function \citep[][]{Naidu:2022,Austin:2023,Donnan:2023,Donnan:2024,Castellano:2023,Castellano2025,Chemerynska2023,Chemerynska:2025,Finkelstein:2024,Adams:2024,Whitler:2025,Atek:2026,McLeod:2026}, all finding that there is a larger than expected abundance of UV-bright galaxies at $z\gtrsim10$. Similarly, the ionizing photon production efficiency of high-redshift galaxies has also been constrained, finding that typical high-redshift galaxies produce a large number of ionizing photons \citep[][]{Endsley:2021,Stefanon:2022,Tang:2023,Simmonds:2023,Simmonds:2024,Harshan:2024,Saxena:2024b,Mascia:2024,Pahl:2025,Hayes:2025}, which have since been mildly revised when large mass-complete samples are used \citep[e.g.,][]{Simmonds:2024b,Austin:2025}. In contrast, the LyC escape fraction is impossible to measure directly in the Epoch of Reionization due to the neutral IGM \citep{Inoue:2014}. Nevertheless, combining measurements of $\rho_{\rm UV}$ and $\xi_{\rm ion}$ with reasonable expectations for $f_{\rm esc}$ of $\sim10\%$ \citep[e.g.,][]{Robertson2015,Rosdahl2022,Jecmen:2026} or extrapolating $f_{\rm esc}$ models based on low-redshift analogues \citep[e.g.,][]{Chisholm2022} suggest that there may be too many ionizing photons: causing the Universe to reionize far earlier than expected based on historic measurements (\citealt{Munoz2024}, though see discussion by \citealt{Papovich:2026}).

A variety of models have since been proposed to explain the observed UV luminosity functions and slow decline in the redshift-evolution of the UV luminosity density beyond redshift $z\sim 10$ (see review by \citealt{Stark:2026}). Popular theories include an increase in the halo-scale star-formation efficiency due to weaker stellar feedback (\citealt{Dekel:2023,Qin:2023,Somerville:2025}, c.f., \citealt{Ferrara:2025}), a decrease in the presence of dust \citep[e.g.,][]{Ferrara:2023}, allowing burstier star formation histories (SFHs) to increase scatter in the $L_{\rm UV}$--$\rm M_{\rm halo}$ relation \citep{Mason:2023,Shen:2023,Sun:2023,Kravtsov:2024}, modifying the star formation rate-UV luminosity relation by including the nebular continuum \citep{Katz2024_BJ}, including a top-heavy IMF \citep[e.g.,][]{Yung:2024,Cueto:2024,Trinca:2024}, modifying the physical properties of stars \citep{Katz2024_He}, and including contributions from AGN \citep[e.g.,][]{Hegde:2024}. 

Due to the fact that it can not be directly detected at high-redshift, the general approach to understand the leakage of ionizing radiation in high-redshift galaxies is to study low-redshift analogue systems \citep[e.g.,][]{Flury:2022,Flury:2022b} and simulations (see review by \citealt{Jaskot:2025}). Once identified, analogue leakers can be used to search for indirect diagnostics: quantities which may be shown to correlate with $f_{\rm esc}$. A huge number of potential diagnostics have been proposed over the last decade, including high $\rm [O~{\small III}]~\lambda5007 / [O~{\small II}]~\lambda\lambda3727$ \citep{Jaskot:2013,Nakajima:2014}, $\rm Ly\alpha$ peak separations \citep{Verhamme:2017,Izotov:2020}, $\rm Ly\alpha$ equivalent width \citep{Steidel:2018,Pahl:2021}, $\rm Ly\alpha$ escape fractions \citep{Dijkstra:2016,Verhamme:2017}, $\rm Mg~{\small II}~\lambda\lambda2796,2804$ doublet ratios \citep{Chisholm:2020,Katz:2022}, $\rm [S~{\small II}]$ deficits \citep{Wang:2021}, strong $\rm C~{\small IV}]~\lambda\lambda1483,1486$ emission \citep{Schaerer:2022,Saxena:2022}, UV slope \citep{Chisholm2022}, $\rm Ly\alpha$-to-$\rm H\alpha$ surface brightness profiles \citep{Choustikov:2024b}, and the $\rm Ly\alpha$ halo fraction \citep{Saldana-Lopez:2025}. Furthermore, multi-variate approaches have also been explored \citep[e.g.][]{Maji:2022,Mascia:2023,Choustikov2024,Jaskot:2024,Jaskot:2024b}, though there is still considerable disagreement. Finally, machine-learning approaches trained on the {\small SPHINX$^{20}$} \citep{Rosdahl2022,Katz:2023} and {\small THESAN-ZOOM} simulations \citep{Kannan2025} to directly predict $\dot{N}_{\rm ion,esc}$ based on \textit{JWST} photometry has also been proposed by \cite{Choustkov:2025} and \cite{Summerfield:2026}. These efforts are complimentary to ongoing attempts to constrain LyC leakage with traditional SED-fitting codes (e.g., \citealt{Giovinazzo:2026,Stoffers:2026,Papovich:2026}, see review by \citealt{Iyer:2026}).

On the theoretical side, considerable effort has gone into simulating the process of LyC escape in both extremely high-resolution GMC-scale simulations \citep[e.g.,][]{Kimm2019,Kimm2022,Kakiichi2021,Menon:2025} and cosmological radiation hydrodynamics simulations \citep[e.g.,][]{Gnedin:2008,Wise2009,Kimm:2014,Paardekooper:2015,Xu:2016,Trebitsch:2017,Trebitsch:2018,Rosdahl2018,Rosdahl2022,Ma:2020,Katz:2020,Katz:2022,Katz2023_bursty,Katz:2025b,Kannan2022,Kannan2025,Yeh:2023,Bhagwat:2024b,Choustikov2024,Choustikov:2024b,Yuan:2024,Giovinazzo:2024,Farcy:2025,Zier:2026}. Here, the difficulty lies in the need to simultaneously resolve the disruption of stellar birth clouds, the multi-phase nature of the interstellar medium (ISM), and the cosmological context of the host galaxy. Of particular relevance here, many potential solutions to the bright galaxy problem rely on modifications to feedback physics -- something which fundamentally modifies the ISM structure in and observable properties of high-redshift galaxies \citep{Choustikov:2025b}.

In this work, we aim to address these challenges by investigating how changes to star-formation/feedback physics inspired by solutions to the bright galaxy problem influence the production and escape of ionizing radiation. To do this, we make use of the suite of high-resolution {\small MEGATRON} simulations \citep{Katz:2025b,Cadiou:2025,Rey:2025b}. These are cosmological radiation hydrodynamics zoom-in simulations of a Milky Way-mass progenitor equipped with a state-of-the-art galaxy formation model. In particular, {\small MEGATRON} features a detailed ISM model coupled through multi-group radiation transport to a 81-species non-equilibrium thermochemical model of primordial species, metals, and molecules -- producing a dynamic and realistic multi-phase ISM for which spectroscopic properties are direct predictions. 

This paper is arranged as follows. In Section \ref{sec:methods}, we describe the {\small MEGATRON} simulations as well as our predictions for LyC escape fractions and emission lines. We discuss the demographics of galaxies that contribute to reionization in Section \ref{sec:reionization}. The ISM structure of LyC leakers is investigated in Section \ref{sec:ISM_structure}, based on which we propose ISM tracers such as electron density as potential diagnostics for the LyC escape fraction in Section \ref{sec:LyC_diagnostics}. Finally, we conclude in Section \ref{sec:conclusions}.

\section{Methods}
\label{sec:methods}
\subsection{The {\small MEGATRON} Simulations}

\noindent For this work, we make use of results from the high-redshift suite of {\small MEGATRON} simulations, described in \cite{Katz:2025b,Katz:2026,Cadiou:2025,Rey:2025b}. These are high-resolution ($\sim\rm  5~pc$ at $z\sim 8.5$) cosmological simulations of a Milky Way progenitor halo simulated using {\small RAMSES-RTZ}\footnote{This is derived from the public versions of {\small RAMSES} \citep{Teyssier2002} and {\small RAMSES-RT} \citep{Rosdahl2013,Rosdahl2015}.} \citep{RTZ}. These simulations include a full non-equilibrium chemistry network including primordial species (H~I-II, He~I-III, and $e^-$), metals (C~I-VI, N~I-VII, O~I-VIII, Ne~I-X, Mg~I-X, Si~I-XI, S~I-XI, and Fe~I-XI), and molecules ($\rm H_2$ and CO) coupled using on-the-fly multi-frequency (8 groups) radiation transport to a detailed model for ISM heating and cooling processes \citep{Katz2022_prism}. As a result, {\small MEGATRON} captures the non-equilibrium state of the multi-phase ISM, which has been shown to reproduce the observed spectroscopic properties of high-redshift galaxies (\citealt{Katz:2026c,Katz:2025b,Choustikov:2025b}, \textcolor{blue}{\textit{in prep.}}), the mass-metallicity relation of local ultra-faint dwarf galaxies \citep{Rey:2025b}, FIR scaling relations (\textcolor{blue}{Matsumoto et al., \textit{in prep.}}),  properties of extreme emission line galaxies (Quilt et al., \textcolor{blue}{\textit{in prep.}}), as well as making predictions for the observability of Population III stars \citep{Storck:2026}, LyC-leaking galaxies (Xing et al., \textcolor{blue}{\textit{in prep.}}), and quenched dwarf galaxies during the Epoch of Reionization \citep{Attard:2026}.

\begin{figure*}
    \centering
    \includegraphics[width=\textwidth]{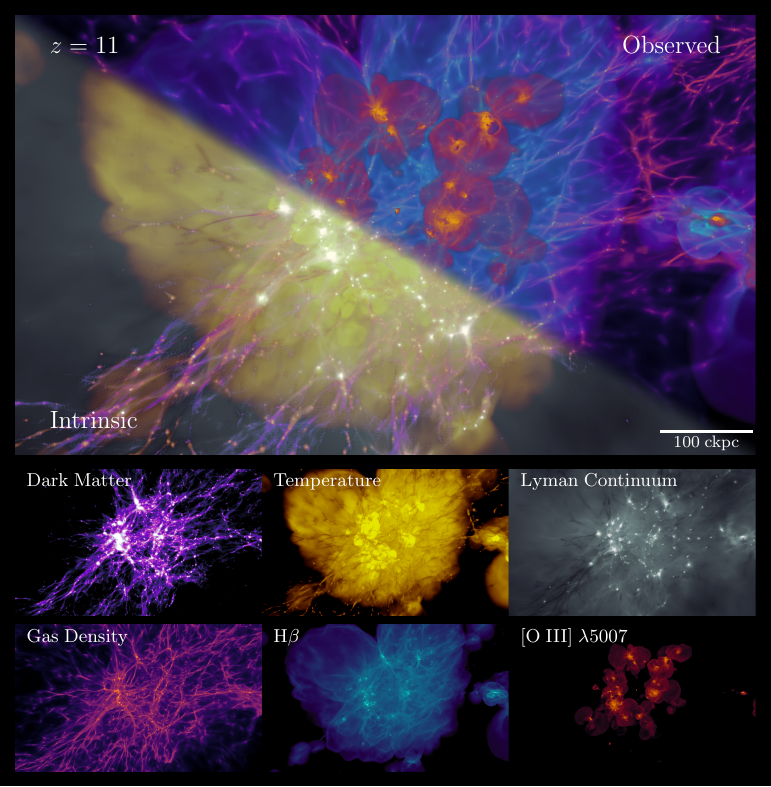}
    \caption{\textbf{An IGM-scale view of the \OG\ {\small MEGATRON} simulation at redshift $\bm{z\sim 11}$.}\\
    A variety of projections through the {\small MEGATRON} simulation are shown. These include intrinsic properties such as the dark matter density field (light purple), gas density (dark purple), temperature (orange), and LyC flux (white), as well as direct predictions of $\rm H\beta$ (blue) and $\rm [O~{\small III}]~\lambda5007$ (red) emission lines. A composite of these features is also shown in the top panel, inheriting the same colours.
    }
    \label{fig:hero}
\end{figure*}

The high-redshift {\small MEGATRON} simulation suite includes four re-simulations of the same initial conditions, equipped with different models for star formation and feedback. These are each designed to capture a different potential solution to the bright galaxy problem and include:

\begin{itemize}
    \item \textbf{\textcolor{OG}{Efficient Star Formation}: }This `fiducial' simulation has the weakest feedback among the four models, leading to the highest global star formation efficiency. Stellar SEDs follow BPASS \citep{Eldridge2017,Stanway2018} assuming a Kroupa IMF with a maximum mass of $\rm 300~M_{\odot}$. This model produces smooth SFHs and galaxies with a very dense ISM along with the widest range of ionization parameters.
    \item \textbf{\textcolor{SF}{Bursty Star Formation}: }This model increases the energy per supernova by a constant factor of five compared to the fiducial run. Galaxy stellar masses are better regulated and the SFHs become significantly burstier. Stellar SEDs follow BPASS \citep{Eldridge2017,Stanway2018} assuming a Kroupa IMF with a maximum mass of $\rm 300~M_{\odot}$.
    \item \textbf{\textcolor{IMF}{Variable IMF}: }This model uses a variable IMF following \cite{Marks2012} with a fixed maximum mass of $\rm 120~M_{\odot}$. As a result, the upper mass slope  becomes shallower at higher gas densities and lower metallicities, leading to a more top-heavy IMF. Some high mass stars explode as hypernovae, releasing $\sim 10\times$ the energy with commensurate yields \citep{Nomoto2006}. Stellar SEDs follow Starburst99 \citep{Leitherer1999}. 
    \item \textbf{\textcolor{ISM}{High $\bm{\epsilon_{\rm ff}}$, HN}: }Stars form in lower-density gas with a local efficiency per free-fall time of $100\%$. A subset of massive stars also explode as hypernovae, as for the variable IMF run. Stellar SEDs follow BPASS \citep{Eldridge2017,Stanway2018} assuming a Kroupa IMF with a maximum mass of $\rm 300~M_{\odot}$.
\end{itemize}

As each model includes different versions of stellar feedback, the typical ISM of a galaxy in each simulation will also be different -- resulting in different emission line ratios and observed properties driven by changes to the density and temperature structure of the ISM \citep[see discussion in][]{Choustikov:2025b}. We make use of these re-simulations to understand the impact of varying star formation/feedback physics on the process and history of reionization. Angle-averaged monochromatic ($\rm 900\AA - 912\AA$) escape fractions are measured at $0.75~R_{\rm vir}$ using \texttt{RASCAS} \citep{Leo2020} as described in \cite{Choustikov:2025b}. 

\subsection{Direct Predictions for Synthetic Observations}

\noindent Across the four simulations, we predict a large library of spectra ($\sim 38,000$ with $\rm M_{\rm UV}\leq-13$ at redshift $z \gtrsim8.5$, see also Figure 4 of \citealt{Katz:2025b}), each accounting for the stellar continuum, nebular continuum and nebular emission lines. For the purpose of this work, we predict the full intrinsic UV-optical spectrum of each galaxy, alongside fluxes for a variety of specific optical lines (e.g., $\rm H\beta$, $\rm [O~{\small II}]~\lambda\lambda 3726,3728$, $\rm [O~{\small III}]~\lambda4363$, and $\rm [O~{\small III}]~\lambda5007$). To do so, we make use of the fact that {\small RAMSES-RTZ} tracks the number density of all relevant ionic species (e.g., $\rm H^+, O^+, O^{2+}$) as well as the electron number density (accounting for ionized metal species) on-the-fly during the simulation. We use {\small PyNeb} \citep{Luridiana2015} to compute the emissivity of each UV/optical emission line on a cell-by-cell basis given the temperature and electron number density\footnote{This includes contributions from all species and metals considered in the simulation.} ($n_e$) of the cell, alongside atomic data from Version 10 of the {\small CHIANTI} database \citep{Dere:1997,Dere2019}. For specific measurements, all gas cells within 25\% of the virial radius of a particular halo are considered\footnote{Convergence tests for this choice of radius are provided in Appendix A of \cite{Choustikov:2025b}.}. This is done in order to avoid the impact of satellite galaxies, though it is important to note that more faithful predictions would require the use of a realistic aperture/slit to account for losses \citep[e.g.,][]{Bhagwat2024}. Finally, we utilise a sub-grid post-processing correction for unresolved Strömgren spheres by using a grid of spherical idealised\footnote{While {\small RAMSES-RTZ} has been shown to reproduce {\small CLOUDY} emission line predictions to within 10\% accuracy in idealised spherically-symmetric setups (see Figure 2 of \citealt{Katz:2026}), this requires a degree of resolution which is beyond that of {\small MEGATRON} under such dense conditions.} {\small CLOUDY} photoionization models \citep{Ferland2017}. The UV slope ($\beta$) is measured by fitting the continuum with a power law of the form $f_{\lambda}\propto\lambda^{\beta}$ within the wavelength range $\rm 1350\AA - 2700\AA$. Following the approach of \cite{Saxena2024}, we include the nebular continuum but exclude nebular emission lines.

\subsection{Measuring ISM Densities}

\noindent In order to be consistent with observational approaches, we also use a variety of strong line ratios to measure the properties of the ISM (see review by \citealt{Kewley2019}). In particular, we focus on the density-sensitive emission line doublet ratio of ${\rm O2} = \rm [O~{\small II}]~\lambda3726/[O~{\small II}]~\lambda3728$ (see Figure 1 of \citealt{Kewley:2019b}). To measure the densities, we again make use of {\small PyNeb} alongside the same atomic data described above. As the combination of temperature-sensitive and density-sensitive lines is fairly uncommon at high redshift (though see discussions in \citealt{Topping:2024b,Arellano-Cordova:2026}), we follow the common procedure of assuming a fixed temperature when measuring densities. Specifically, we adopt a fiducial temperature of $T_{\rm e}=1.5\times10^4~{\rm K}$\footnote{This adoption was made to be consistent with high-redshift observational studies that typically assume densities in the range $10\,000 - 20\,000~\rm K$ \citep[e.g.,][]{Isobe2023,Abdurro'uf:2024,Li:2025,Topping2025,Stephenson:2026}. Changes to this assumption only begin to affect measurements at $n_e\lesssim10~\rm cm^{-3}$.}. 

Throughout the present work, we frequently make use of two different density measurements, defined as follows. The `intrinsic' gas density ($\langle n_{\rm H} \rangle_{\rm [OII]}$) is the average hydrogen density weighed by the luminosity of $\rm [O~{\small II}]~\lambda\lambda3727$ as computed within the ISM of each galaxy. In contrast, the `observed' electron density ($n_e([{\rm OII}])$) is the inferred density as measured from the $\rm [O~{\small II}]$ doublet as described above. While the former is not directly observable, it is a useful feature to consider in a theoretical context as it is not affected by any observational bias. A comparison between these two quantities is provided in Appendix \ref{app:nH_ne}.

\subsection{Ionizing Photon Production Efficiency}

\noindent To compute ionizing photon production efficiencies ($\xi_{\rm ion}$), we follow the typical approach of assuming Case~B recombination with $T_{\rm e}=1.5\times10^4~{\rm K}$, $n_e=100~{\rm cm}^{-3}$, $f_{\rm esc} = 0$ so that \citep{Osterbrock2006}:
\begin{equation}
    \xi_{\rm ion}({\rm H\alpha})\;/\;[{\rm Hz}/{\rm erg}] = \frac{7.7\times 10^{11} L({\rm H\alpha})}{L_{1500,\rm int}}\,
\end{equation}
where $L(\rm H\alpha)$ is the flux of the $\rm H\alpha$ line and $L_{1500,\rm int}$ is the intrinsic UV luminosity density \citep[in the $L_{\nu}$ system, ][]{Maseda:2020}. This efficiency should in principle be corrected for the escape fraction of LyC radiation\footnote{Such measurements also ignore the presence of nebular LyC emission, see discussion by \cite{Simmonds:2024c}.}. However, to best compare with observational inferences, we ignore this effect (see discussion in \citealt{Saxena:2024}). The intrinsic ionizing photon production efficiency, $\xi_{\rm ion}^{\rm int}$, is instead computed using the intrinsic ionizing luminosity of all stars within the central $0.25R_{\rm vir}$ of the halo in question.

\section{The Sources of Reionization in MEGATRON}
\label{sec:reionization}

\noindent Our goal is to understand the sources and history of reionization in the high-redshift {\small MEGATRON} simulations. Figure \ref{fig:hero} shows a compilation of projections through the cosmic web and IGM of the \OG\ simulation at redshift $z = 11$, showing the appearance of an ionized bubble centreed on the dark matter overdensity, driven by a large number of sources. In particular, we show the dark matter density, gas density, mass-weighed temperature and LyC photon flux, as well as predictions for $\rm H\beta$ and $\rm [O~{\small III}]~\lambda 5007$ luminosities. At this stage, the reionization process has only just begun ($x_{\rm HII}\lesssim0.1$, see below). Furthermore, metal enrichment has not significantly polluted the IGM and oxygen emission is confined to star-forming galaxies. Finally, while some strong LyC sources appear co-spatial with strong emission lines, there are also some very strong sources that do not, consistent with a scenario where all line-emitting gas has been evacuated.

\begin{figure}
    \centering
    \includegraphics[]{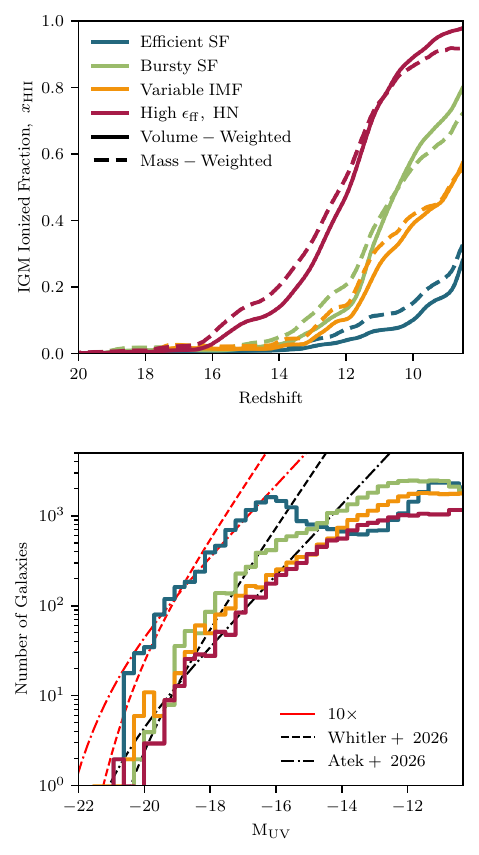}
    \caption{\textbf{Reionization history and UV number counts.}\\
    Redshift evolution of the IGM ionized fraction ($x_{\rm HII}$, top) till the end of the simulation at $z\sim 8.5$ as well as the measured distribution of UV magnitudes (bottom) among snapshots below redshift $z\lesssim 16$ for each {\small MEGATRON simulation}. We include observed UVLFs renormalised to the \ISM\ simulation at $\rm M_{\rm UV} = -19$ for comparison. While the \ISM\ simulation features the fewest galaxies, it produces the fastest reionization history. The opposite is true for the \OG\ simulation.}
    \label{fig:reionization_history}
\end{figure}

\begin{figure}
   \centering
   \includegraphics[]{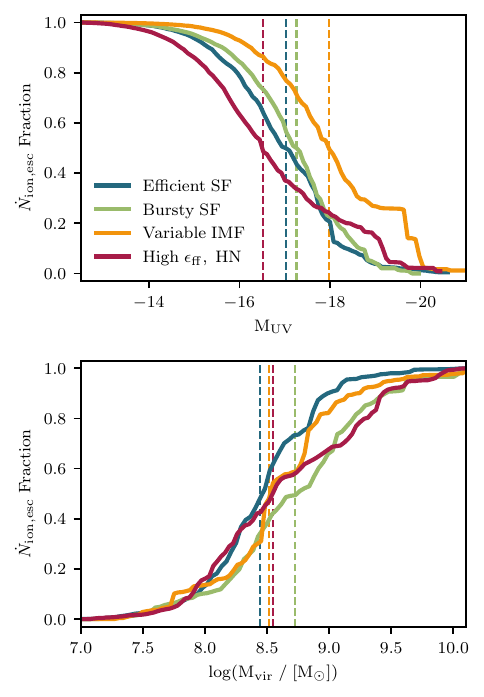}
   \caption{\textbf{Galactic contributions to Reionization.}\\
   Cumulative fractions of the total escaping LyC photons released by galaxies below redshift $z\leq14$ as a function of UV magnitude (top) and halo virial mass (bottom) for each {\small MEGATRON} simulation. Magnitudes and masses corresponding to $50\$$ of the contributed LyC photons are denoted by dashed vertical lines. Different feedback models lead to changes in the relative importance of UV bright/faint galaxies, while leaving the dependence on halo mass less impacted.}
   \label{fig:source_cdfs}
\end{figure}

The evolution of the mass- and volume-weighted IGM ionized fraction, $x_{\rm HII}$ as computed within the entire high-resolution zoom region is shown in the top panel of Figure \ref{fig:reionization_history} (see also Appendix C of \citealt{Rey:2025b}). We find that the four different star formation/feedback prescriptions lead to dramatically different reionization histories. For example, this process is almost entirely complete in the \ISM\ simulation by redshift $z\sim 8.5$, while only a third of the \OG\ simulation volume has been ionized at this time. These differences are interesting when contrasted against the distribution of source UV magnitudes for each simulation at redshifts $z\lesssim16$ (bottom panel of Figure \ref{fig:reionization_history}). Here, we find that the \OG\ simulation features by far the most (by an order of magnitude) UV-bright ($\rm M_{\rm UV}\leq -18$) sources, in comparison to the \ISM\ simulation which features the smallest number of galaxies at nearly any $\rm M_{\rm UV}$. We compare each simulation to observed UV luminosity functions (UVLFs) from \cite{Whitler:2025} and \cite{Atek:2026}, each normalised\footnote{As the {\small MEGATRON} suite are all cosmological zoom-in simulations of a biased region \citep{Katz:2025b,Rey:2025b}, it is difficult to fairly test the normalisation of the predicted UVLF to observed samples without making extra assumptions \citep[see discussion in][]{Ma2018}. Therefore, we restrict comparisons to the shape of the observed UVLF.} to the number of galaxies in the \ISM\ simulation at $\rm M_{\rm UV}=-19$. Interestingly, the \OG\ and \ISM\ simulations both prefer the shape of the flatter faint-end slope of \cite{Atek:2026}. In contrast, the \SF\ simulation prefers the steeper slope of \cite{Whitler:2025}, while the \IMF\ simulation falls somewhere in the middle. These findings align with previous simulations have shown that modifications to feedback physics such as stellar SEDs \citep{Rosdahl2018}, IMF \citep{Cueto:2024}, stellar feedback strength/timing \citep{Bhagwat:2024b}, and cosmic rays (\citealt{Farcy:2025}, see also discussion by \citealt{Martin-Alvarez:2026}) can jointly modify predicted UVLFs alongside the process of reionization. Nevertheless, it is instructive to understand the physical reasons for these differences, especially given the naive expectation that a greater number of UV sources would lead to a faster reionization history.

\subsection{Demographics of LyC Leakers in {\small MEGATRON}}

\begin{figure*}
   \centering
   \includegraphics[width=\textwidth]{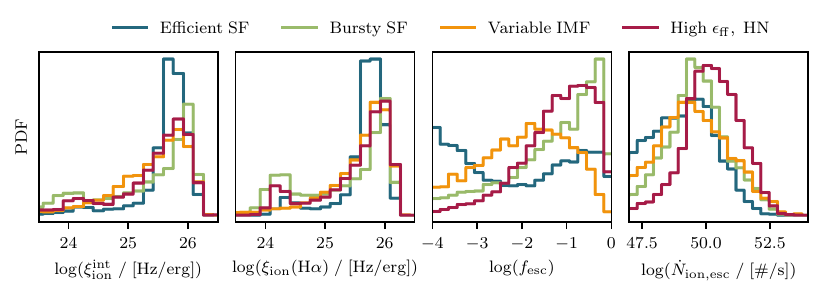}
   \caption{\textbf{Escape fraction, ionising production efficiency, and escaping ionising luminosity population demographics.}\\
   We show distributions for $\xi_{\rm ion}$ (both the intrinsic value alongside that inferred from $\rm H\alpha$), $f_{\rm esc}$, and $\dot{N}_{\rm ion,esc}$ for all galaxies with $\rm M_{\rm UV}\leq-13$ across all four {\small MEGATRON} simulations. Changes to star formation/feedback physics modify escape fraction distributions while leaving $\xi_{\rm ion}({\rm H\alpha})$ largely unchanged. However, the mapping to $\dot{N}_{\rm ion,esc}$ is non-trivial. For example, while the \SF\ model produces the largest number density of strong leakers ($f_{\rm esc}\gtrsim30\%$), the ionizing output of these galaxies is systematically lower than those of the \ISM\ simulation.}
   \label{fig:fesc_xi_nion_hist}
\end{figure*}

\begin{figure}
   \centering
   \includegraphics[]{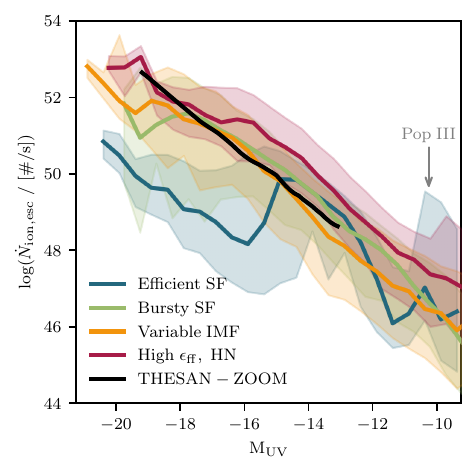}
   \caption{\textbf{Ionizing Luminosity as a function of Magnitude.}\\
   We show running medians for the escaping ionizing luminosity of galaxies as a function of $\rm M_{\rm UV}$ for galaxies below redshift $z\leq16$ across all four {\small MEGATRON} simulations. Both the normalisation and scatter (shown by shaded areas which mark the $16^{\rm th}$ and $84^{\rm th}$ percentiles) in this relation are sensitive to feedback physics and can be conditional on $\rm M_{\rm UV}$. Among very faint galaxies ($\rm M_{\rm UV}\sim -10$), there is a boost due to the ionizing luminosity of Population III stars. As a comparison, we show the running median from the {\small THESAN-ZOOM} simulation \protect\citep{Summerfield:2026}. }
   \label{fig:Nion_muv}
\end{figure}

\noindent Due to the fact that the process of reionization depends on the ionizing emissivity, an accelerated reionization history can canonically be attained by either increasing the number of sources, the ionizing photon production efficiency, or the escape fraction. In reality however, these three parameters are intrinsically linked by the processes of star formation and feedback -- something that we now aim to disentangle. As the {\small MEGATRON} simulations follow a zoomed-in unrepresentative region of the Universe, we can not reliably report on traditional volume-averaged quantities (see discussions in \citealt{Lovell2021,Lovell:2026}). Instead, we leverage the fact that each simulation uses identical initial conditions and therefore all feature the same controlled volume -- allowing us to make meaningful benchmark comparisons.

To understand the demographics of LyC sources, Figure \ref{fig:source_cdfs} shows cumulative distribution functions (CDFs) of the escaping ionising luminosity ($\dot{N}_{\rm ion,esc}$) for each simulation as a function of UV magnitude (top) and halo mass (bottom). Magnitudes and masses corresponding to 50\% of the contributed LyC photons are denoted by dashed vertical lines. We find that the LyC budget in the \IMF\ simulation is dominated by UV-bright galaxies (due to the different $\rm M_{\rm UV}-M_*$ relation, see Appendix \ref{app:scaling_relations}), in contrast to the \ISM\ simulation where $\sim 40\%$ of LyC photons are produced by galaxies with $\rm M_{\rm UV}\leq -16$ (see discussion below). In contrast, we find that the relative contributions of LyC-contributing DM halos are less sensitive to feedback, owing to the fact that the median $\dot{N}_{\rm ion,esc}-{\rm M}_{\rm vir}$ appears comparably robust (albeit with more scatter than in other relations, see also Appendix \ref{app:scaling_relations} for a discussion of scaling relations). This conclusion brings a novel angle to the ongoing debate of whether reionization is a democratic process or driven by a small number of massive, UV-bright halos \citep[e.g.,][]{Finkelstein2019,Naidu:2020}, by highlighting that the $\rm M_{\rm UV}-M_{\rm vir}$ relation is linked to the physics of LyC escape.

Figure \ref{fig:fesc_xi_nion_hist} shows the distributions of intrinsic and $\rm H\alpha$-inferred $\xi_{\rm ion}$, $f_{\rm esc}$, and $\dot{N}_{\rm ion,esc}$ for each simulation, considering all galaxies with $\rm M_{\rm UV}\leq-13$. It is important to note that while Population III-dominated galaxies are included in this selection, these are unlikely to dominate due to the low UV luminosity of such systems (see discussion in \citealt{Storck:2026}). We find that all four simulations produce similar ionizing photon production efficiencies (up to differences in SEDs used and changes to the density structure of the ISM, see also \citealt{Katz2024_BJ} for discussions of impacts due to the nebular continuum). However, there is a slight bias in that the strongest intrinsic producing galaxies ($\xi_{\rm ion}\gtrsim10^{26}~\rm Hz/erg$) appear to be comparatively less common in the \IMF\ and \ISM\ simulations when inferred by Balmer emission, as compared to the frequency of such sources in the \OG\ and \SF\ simulations. This is attributed to differences in the nebular continuum driven by whether or not the ISM has experienced feedback from HNe as well as the impact of a non-zero ionizing escape fraction. Furthermore, we find markedly different distributions in $f_{\rm esc}$ as well as $\dot{N}_{\rm ion,esc}$. This demonstrates that the implementation of feedback has a significant effect on the creation of under-dense ionized channels through which LyC photons can escape \citep[see also discussions in][]{Kimm2022,Bhagwat:2024b,Farcy:2025}. In particular, we find that the \SF\ simulation produces the largest number density of galaxies with $f_{\rm esc}\gtrsim 10\%$, followed by the \ISM\ and \OG\ simulations. Interestingly, the \IMF\ simulation features by far the fewest number density of galaxies with $f_{\rm esc}\gtrsim 10\%$.

To synthesize these findings, Figure \ref{fig:Nion_muv} shows the median $\dot{N}_{\rm ion,esc}$ as a function of $\rm M_{\rm UV}$ for galaxies below redshift $z\leq16$ across all four {\small MEGATRON} simulations. We find that both the normalisation and scatter (shaded regions describe the $16^{\rm th}-84^{\rm th}$ percentiles) are sensitive to feedback physics. In particular, while the \SF\ and \IMF\ simulations follow a common trend (which agrees with findings from {\small THESAN-ZOOM}, \citealt{Summerfield:2026}), there is a marked step change in behaviour around $\rm M_{\rm UV}\sim-16$ in the other two simulations. Namely, galaxies fainter than this limit in the \ISM\ simulation contribute $\sim 1~\rm dex$ more ionizing radiation, while galaxies brighter than this in the \OG\ simulation release $\sim 2~\rm dex$ fewer ionizing photons than comparable systems -- mirroring the opposite of the observed behaviour in the distribution of UV sources for each simulation (Figure \ref{fig:reionization_history}). Nevertheless, galaxies brighter than $\rm M_{\rm UV}\leq-17$ in the \OG\ simulation still account for $50\%$ of the total ionizing background due to their increased abundance. Interestingly, there is also a noticeable up-scatter in $\dot{N}_{\rm ion,esc}$ at $\rm M_{\rm UV}\sim -10$ due to the presence of Population III stars, which release more ionizing photons per stellar mass \citep{Schaerer:2002,Larkin2023}.

Finally, it is crucial to highlight that the distribution of LyC escape fractions does not translate to effective contributions to reionization. This is due to the existence of remnant leakers (galaxies with large escape fractions but no LyC-producing stars, \citealt{Katz2023_bursty}). Similarly, models which are capable of effectively quenching dwarf galaxies by expelling all gas will produce more remnant leakers, stopping such galaxies from contributing to reionization. In particular, \cite{Attard:2026} recently showed that the \SF\ and \ISM\ simulations produce significantly larger fractions of quenched galaxies. As a result, while the \SF\ simulation produces significantly more galaxies with $f_{\rm esc}>10\%$ than the \IMF\ simulation, both simulations produce the same number density of galaxies with $\dot{N}_{\rm ion,esc}\gtrsim10^{51}~{\rm s^{-1}}$. Ultimately, these differences are driven by the timescale synchronicity of LyC production and leakage.

\subsection{The Importance of Timing for Effective Leakage of LyC Photons}

\noindent The question of LyC escape from galaxies is one of timing. Immediately following a burst of star formation, the race is on to clear channels in the ISM before the sources of ionising radiation disappear. In the extreme case, we can understand this by considering a stellar population consisting of a single O star. Without early feedback\footnote{Strong turbulence is also capable of creating sightlines with non-zero escape fractions, owing to the broadening of the gas density PDF and the probabilistic production of low-column density channels.}, the first chance that LyC photons will have to escape will come after the star goes supernova -- at which point there will necessarily be no more photons to escape. In this scenario, there will be a step change at $\sim 3~{\rm Myr}$, at which point the escape fraction increases significantly while the ionising luminosity reaches $\dot{N}_{\rm ion,esc} = 0$. If instead we consider early feedback (e.g., winds, radiation pressure, photonionization) then this transition will be smoothed -- allowing the O star to contribute. In reality, we must contend with multiple concurrent sources of both feedback and LyC photons, as well as the necessity to clear channels beyond the birth cloud.

\begin{figure}
   \centering
   \includegraphics[]{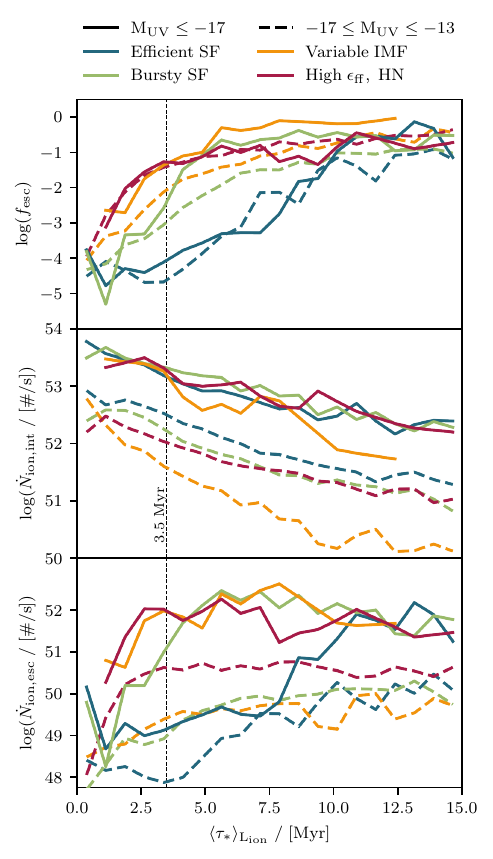}
   \caption{\textbf{Time-evolution of LyC production and escape.}\\
   We show running medians of $f_{\rm esc}$, $\dot{N}_{\rm ion,int}$, and $\dot{N}_{\rm ion,esc}$ as a function of the ionizing luminosity-weighted mean stellar age for galaxies in each {\small MEGATRON} simulation with $\rm M_{\rm UV}\leq-17$ (solid) and $-17 \leq\rm M_{\rm UV}\leq-13$ (dashed). Changes to feedback physics (e.g., burstier star formation, stronger early feedback, and preferentially forming stars in low-density gas) all act to shorten the time delay between a particular starburst and the onset of effective leakage. As the intrinsic ionizing luminosity decreases with time, a shorter gap will lead to more effective leakage. To highlight this, we include a dashed vertical line at $\sim 3.5~{\rm Myr}$, at which point the first Type II SNe begin. Interestingly, the UV-bright population is capable of releasing similar numbers of ionizing photons, while the output of UV-faint galaxies is strongly dependent on feedback.}
   \label{fig:time_evo}
\end{figure}

To understand the timescales for effective leakage produced in a realistic context, Figure \ref{fig:time_evo} shows running medians of $f_{\rm esc}$, the intrinsic ionising luminosity ($\dot{N}_{\rm ion,int}$) and $\dot{N}_{\rm ion,esc}$ for galaxies with $\rm -13 \geq M_{\rm UV}\geq -17$ (dashed) and $\rm M_{\rm UV}\leq -17$ (solid) in each simulation as a function of ionising luminosity-weighted mean stellar age\footnote{This was done due to the fact that simulation snapshots were saved at a coarser cadence ($\sim 2-5~\rm Myr$) than is necessary to capture physical processes relevant to the present discussion.}. We find that simulations that feature hypernovae begin to leak more quickly \citep{Nomoto2006} and sustain a higher level of leakage for all subsequent times. In particular, the \ISM\ simulation features the shortest time delay between burst and leakage. This is driven by the fact that stars form in lower-density gas in this simulation, leading to more effective coupling of radiation and SNe feedback to the ISM (see discussions in \citealt{Blondin:1998,Thornton:1998,Kannan:2020}). Following this, the median galaxy in the \SF\ simulation also shows boosted leakage at earlier times, driven by the boosted SNe energy \citep{Katz:2025b}. The median \OG\ simulation galaxy only reaches the highest escape fractions after $\sim 10~{\rm Myr}$. As highlighted by \cite{Katz2023_bursty} and \cite{Choustikov2024,Choustikov:2024b}, high levels of leakage are inconsequential if stellar populations are not producing any ionising photons. Overall, this means that burstier star formation, stronger early feedback, modifications to stellar physics which lead to star formation in lower density environments or the inclusion of hypernovae will all lead to earlier leakage as well as higher escape fractions at all times. Finally, it is interesting to note that while UV-bright ($\rm M_{\rm UV}\leq-17$) galaxies in the \SF\ and \IMF\ simulations begin leaking earlier than UV-faint ($-17\leq\rm M_{\rm UV}\leq-13$) systems, owing to the high gas densities of UV-faint systems undergoing their first major burst of star formation. In contrasts, the trends discussed above are unchanged with luminosity for the \OG\ and \ISM\ models, as they either all have very high gas densities, or low gas densities respectively.

In the central panel of Figure \ref{fig:time_evo}, we can see that on average, the ionising luminosity of galaxies in all simulations decays as a function of their age, especially after $\sim 3~{\rm Myr}$ at which point the first O stars start to die \citep[][]{Yusof:2013,Groh:2014,Chisholm:2019}. However, it is important to note that we find two key patterns of behaviour. First, galaxies in the \IMF\ simulation tend to produce fewer ionising photons at later ages. This is due to the fact that this simulation uses Starburst99 SEDs \citep{Leitherer1999} (as opposed to the others, which use BPASS, \citealt{Eldridge2017,Stanway2018}), which do not consider binary stars and limit the maximum mass to 120$~\rm M_{\odot}$. While a top-heavy IMF will help to bridge this gap by increasing the fraction of massive stars in young stellar populations, binarity acts to increase the magnitude and duration of LyC production \citep{Stanway:2016,Gotberg:2017,Gotberg:2020} and are known to have a significant impact on simulations of reionization \citep{Ma2016,Rosdahl2018}. Furthermore, very massive stars can also be an important factor by significantly boosting the ionizing output and early feedback\footnote{It is important to note that stellar winds can be significantly weaker at low-metallicity, leading to delayed feedback (see discussion in \citealt{Jecmen2023}).} of a given stellar population \citep[e.g.,][]{Schaerer:2025}. Next, we find that UV-faint galaxies in the \ISM\ simulation also produces systematically fewer ionising photons at early times, owing to the inclusion of hypernovae physics, allowing stars with masses beyond $25~\rm M_{\odot}$ to explode much earlier (see also \citealt{Kobayashi:2007}). In contrast, UV-faint galaxies in the \OG\ simulation tend to produce more ionising photons at all ages. This is due to the fact that smoother SFHs (see e.g., Figure 5 of \citealt{Katz:2025b}) lead to multiple concurrent starbursts with more continuous LyC production. At this stage, it is relevant to note that an ionising luminosity-weighted stellar age is not a strict measure of time since starburst in galaxies with extended SFHs (see also discussion in \citealt{Mehrotra:2026}). To this end, we shall discuss an alternate probe below.

\begin{figure}
   \centering
   \includegraphics[]{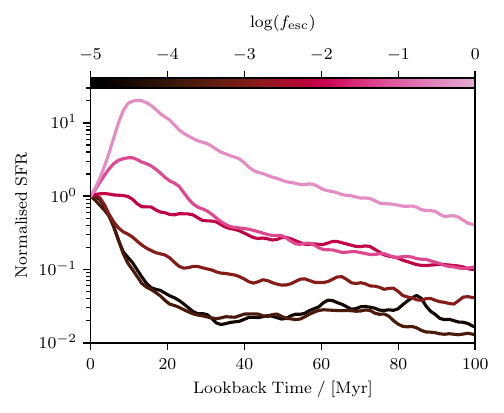}
   \includegraphics[]{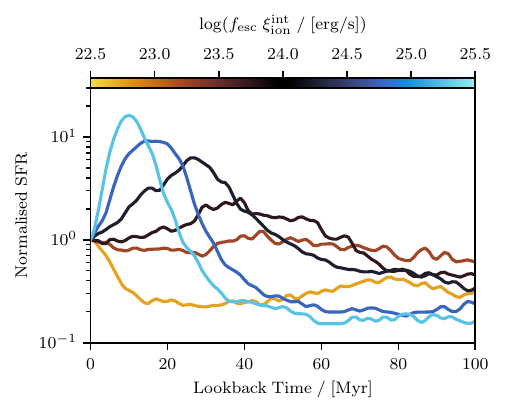}
   \caption{\textbf{Typical SFHs of LyC-leaking galaxies.}\\
   We show median SFHs of galaxies with $\rm M_{\rm UV}\leq-15$ in the \SF\ simulation, stacked as a function of $f_{\rm esc}$ (top) and $f_{\rm esc}\xi_{\rm ion}^{\rm int}$ (bottom). These have been smoothed with a Gaussian kernel of standard deviation $\sim 1~\rm Myr$ and renormalised at the present day. Rapidly rising SFHs are emblematic of non-leakers, while on average, LyC-leaking systems appear to be in the post-starburst phase due to the need for a strong feedback episode to clear channels through the neutral ISM.}
   \label{fig:sfr_stacks}
\end{figure}

Finally, the bottom panel of Figure \ref{fig:time_evo} shows the median $\dot{N}_{\rm ion,esc}$ as a function of ionising luminosity-weighted stellar age in each simulation. Here, we find that changes to feedback physics affect galaxies differently as a function of their UV magnitude. Among the UV-bright systems ($\rm M_{\rm UV}\leq-17$), all four models converge to similar escaping ionizing luminosities ($\dot{N}_{\rm ion,esc}\sim10^{52}~\rm s^{-1}$), though after very different time delays. Both simulations  featuring HNe reach this stage after $\sim 2.5~{\rm Myr}$, while the \SF\ and \OG\ simulations begin leaking at this rate after $\sim 4~{\rm Myr}$ and $\sim 11~{\rm Myr}$ respectively. In contrast, UV-faint ($-17\leq\rm M_{\rm UV}\leq-13$) galaxies in the \ISM\ simulation release more ionising photons at all times, in particular $\sim 3.5~{\rm Myr}$ into the bursty star formation cycle, as indicated by the vertical dotted line. As a direct consequence, this simulation begins to reionize the IGM significantly earlier than the others, despite the fact that it features fewer galaxies at a given $\rm M_{\rm UV}$ (see Figure \ref{fig:reionization_history}). Next, we find that typical UV-faint galaxies in the \IMF\ and \SF\ simulations follow very similar trends, though the former has slightly more leakage before $\sim 3.5~{\rm Myr}$, before the latter begins to dominate at later times due to the presence of binary stars which increase the production efficiency at older stellar populations. Finally, we find that the time evolution of $\dot{N}_{\rm ion, esc}$ is fundamentally different in the \OG\ simulation. This is due to the fact that feedback is weaker, meaning that galaxies in this simulation can over-cool -- requiring multiple successive starbursts to clear a particularly dense ISM (see also discussion in \citealt{Choustikov:2025b}).

It is important to note that the ionizing luminosity-weighted stellar age is sensitive to the SFH, particularly for galaxies with complex or extended SFHs, and may therefore not directly correspond to the time since the last starburst. Therefore, to confirm our findings, Figure \ref{fig:sfr_stacks} shows a compilation of smoothed (with a standard deviation of $1~\rm Myr$) median SFHs computed for stacks in $f_{\rm esc}$ (top) and $f_{\rm esc}\xi_{\rm ion}^{\rm int}$ (bottom) considering all galaxies with $\rm M_{\rm UV}\leq-15$ in the \SF\ simulation. Regions where the median SFH appears flat is caused by the fact that we are averaging over many galaxies which would have experienced bursts at different times. In both cases, we find that galaxies which can and do leak more ionizing photons tend to show a post starburst feature, in contrast to non-leakers which feature rapidly rising SFHs representative of an ongoing burst of star formation. These results echo the trends highlighted for the \SF\ simulation in Figure \ref{fig:time_evo}: that there is often a characteristic time delay between the onset of a starburst and the conditions necessary for effective leakage. Furthermore, we note that this time delay is similar to that of the `bursty' leakers found in the {\small SPHINX$^{20}$} simulation (see Figure 3 of \citealt{Katz2023_bursty}).

\begin{figure*}
    \centering
    \includegraphics[width=\textwidth]{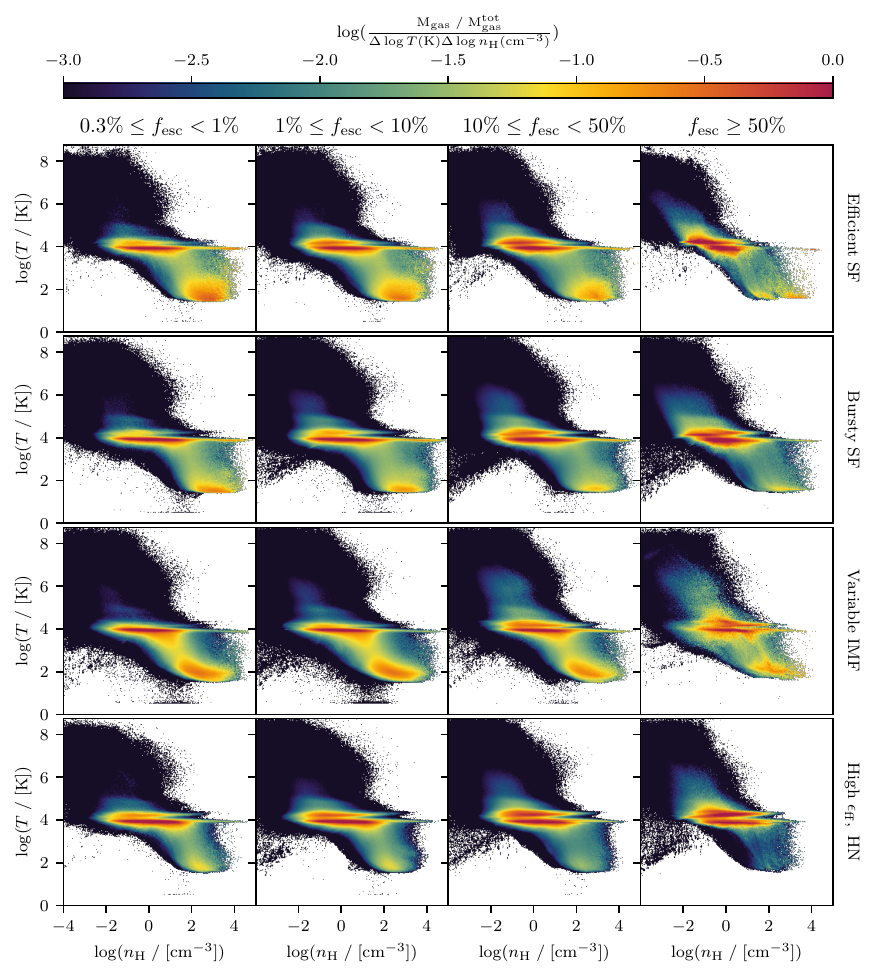}
    \caption{\textbf{Multi-phase ISM structure of non-leaking galaxies compared to intermediate and strong LyC leakers.}\\
    In each panel, we show mass-weighed density-temperature phase diagrams of all gas in the ISM of a mass-matched ($10^6\le {\rm M_*/M_{\odot}\le 10^{6.5}}$) sample of galaxies. Each column corresponds to a stack of galaxies with LyC escape fractions within a specific range (larger $f_{\rm esc}$ from left to right). Similarly, each row corresponds to the relevant stack for each {\small MEGATRON} simulation. While the intrinsic ISM structure is different for each simulation as a function of changes to star-formation/feedback physics, the mechanisms that drive the production of escape channels are unchanged. Strong photoionization feedback is present in intermediate leakers ($1\%\le f_{\rm esc}\le10\%$), strong leakers ($10\%\le f_{\rm esc}\le50\%$) display a combination of photoionization and SNe feedback, while the cold/dense ISM of extreme leakers ($f_{\rm esc}\ge50\%$) is almost completely disrupted.}
    \label{fig:full_phase_diagrams}
\end{figure*}

\subsection{Sensitivity of LyC Leakage Timescales to\\Feedback Physics}

\noindent Interestingly, we have found that the presence of earlier feedback alongside the drive to form stars in lower-density gas (as used in the \ISM\ simulation) means that effective leakage can start earlier than the $\sim 3.5~{\rm Myr}$ timescale discussed in \cite{Choustikov2024,Choustikov:2024b}, shown by a dotted line in Figure \ref{fig:time_evo}. In contrast, the average galaxy in the other simulations begins leaking at or after this timescale, potentially driven by the time taken to photoionize and/or clear channels through the more dense ISM produced by these models (\citealt{Choustikov:2025b}, \textcolor{blue}{\textit{in prep}.}). While numerous studies based around cosmological simulations have shown that there are characteristic feedback-dependent delays between starbursts and high escape fractions \citep[e.g.,][]{Paardekooper:2015,Trebitsch2017,Ma:2020,Barrow:2020,Katz2023_bursty,Choustikov2024}, the exact timescale is not settled. For example, the simulations of \cite{Kimm:2014} found a characteristic leakage timescale of $\sim10~\rm Myr$, equal to the mean main-sequence lifetime of a SNe progenitor assumed therein \citep{Schaller:1992}.

Furthermore, the question of whether SNe are the key feedback mechanism is also debated. For example, high resolution simulations of GMC-scale simulations \citep[e.g.,][]{Kim:2019,Kimm2019,Kimm2022} seemingly agree that early stellar feedback is more important, particularly focusing on the impact of photoionization heating (see Section 4.1 of \citealt{Rosdahl2022} for a discussion of this physics in {\small SPHINX$^{20}$}). Furthermore, \cite{Menon:2025} find that the timescale to clear the birth cloud is shorter for star-forming regions with higher gas surface densities, owing to the shortening of collapse and gas consumption timescales alongside the added impact of radiation pressure in such high-density environments. While this is physics that operates on resolutions that are not always resolved in {\small MEGATRON}, it is important to note that while GMC-scale simulations follow the star-forming environment with exquisite detail, they do not include the bulk ISM of the host galaxy and therefore can only speak to the feedback necessary for radiation to escape past the birth cloud (see Figures 8 and 9 of \citealt{Yoo:2020}).

At the same time, several recent low-redshift observations have pointed out that direct radiation pressure may be the key feedback process to induce leakage, with observed LyC emitters featuring ages indicative of pre-SNe stellar populations \citep[e.g.,][]{Komarova2025,Carr:2025b,Carr2025}. However, it is possible that this is a temporary line-of-sight effect driven by the lack of LyC production following the death of massive O stars, with other observations suggesting that SNe-driven outflows can provide effective escape channels for neighbouring stellar populations \citep{Peng:2023,Martin:2024}. Nevertheless, it has been well established that the pre-processing effect of radiation feedback is vital \citep[e.g.,][]{Rosdahl:2015,Petersson:2025} and further observations to better constrain the impact of early stellar feedback on angle-averaged $f_{\rm esc}$ are needed. Finally, the efficiency of pre-SNe feedback in the context of a dense ISM characteristic to starburst galaxies in the early Universe \citep{Isobe2023,Reddy:2023,Abdurro'uf:2024,Senchyna2024,Topping:2024b,Topping2025,Li:2025} remains to be seen, where physics not presently included in {\small MEGATRON} such as $\rm Ly\alpha$ pressure \citep[e.g.,][]{Smith:2016,Kimm2018,Nebrin2025,Nebrin:2026,Menon:2026} and cosmic rays \citep[e.g.,][]{RodriguezMontero:2024,Farcy:2025} may also be vital pieces of the puzzle .

Altogether, we find that more effective, earlier feedback leads to a scenario where a smaller number of galaxies (across all UV magnitudes) are capable of driving significantly earlier reionization. In contrast, smoother SFHs with less effective feedback lead to a much larger number of bright sources but a much slower reionization process owing to the delay before effective leakage in individual galaxies. Furthermore, feedback in lower-density gaseous environments will also lead to escape of ionizing radiation earlier into the burst cycle. However, irrespective of feedback, the DM halos that typically contribute to reionization remain unchanged.

\section{The ISM Structure of UV-faint LyC Leakers in MEGATRON}
\label{sec:ISM_structure}

\begin{figure}
    \hspace{-0.5cm}
    \includegraphics[width=0.52\textwidth]{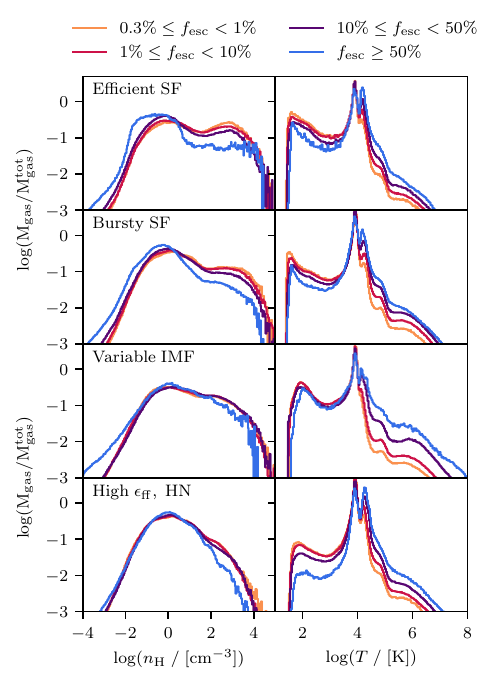}
    \caption{\textbf{Density/temperature structure of LyC leakers.}\\
    We show ISM gas mass fractions as a function of density (left) and temperature (right) for each {\small MEGATRON} simulation (rows). Each line is computed by integrating the relevant phase diagram of Figure \ref{fig:full_phase_diagrams}. The ISM of galaxies with larger LyC escape fractions feature progressively less dense/cold gas and significantly more hot gas. Similarly, the ratio of peaks at $\sim 10^4~{\rm K}$ and $\sim 10^{4.5} ~{\rm K}$ is sensitive to $f_{\rm esc}$, indicating the presence of strong photoionization feedback.}
    \label{fig:integrated_phase_diagrams}
\end{figure}

\noindent Across all four {\small MEGATRON} simulations, we find that UV-faint, low mass DM halos can contribute significantly to reionization of their local environment. Therefore, it is instructive to understand the physics that drives leakage in such systems.

To accomplish this, we investigate the structure of the multi-phase ISM, making use of the detailed PRISM ISM model used in {\small MEGATRON} \citep{Katz2022_prism,Katz:2026}. For each simulation, we compute mass-weighted density-temperature phase diagrams for a mass-matched sample (see e.g., \citealt{Kim:2017,Katz2022_prism} for a discussion). Specifically, we consider all gas within the ISM ($\rm R<0.25~R_{\rm vir}$) of all galaxies with $10^6\leq \rm M_*/M_{\odot}\leq10^{6.5}$, before then stacking these based on LyC escape fractions of each halo. Figure \ref{fig:full_phase_diagrams} shows the resulting phase diagrams, organised such that each column denotes a different range of $f_{\rm esc}$ (stronger leakers on the right), with each simulation shown on a unique row. Focusing first on the left-most column, we confirm that while each {\small MEGATRON} simulation produces a multi-phase ISM, the exact structure and properties of the ISM are different in each simulation, owing to the different implementations of feedback and leading directly to differences in their observed properties \citep{Katz:2026c,Katz:2025b,Katz:2026,Choustikov:2025b,Attard:2026}.

Comparing the different $f_{\rm esc}$ bins for the \OG\ simulation (top row), we find that galaxies with $1\%\leq f_{\rm esc}\leq10\%$ show significantly more gas at $15,000~{\rm K}-20,000~{\rm K}$, as compared to the weakest leakers considered, where the majority of photoionized gas sits at $10,000~{\rm K}$\footnote{Not all of the gas at $10,000~{\rm K}$ is photoionized, with a large part constituting the warm neutral medium (WNM), see discussions in \cite{Kim:2023b}.} \citep[e.g.,][]{Rosdahl:2015}. The difference here amounts to whether the photoionized gas is mixed with the neutral ISM, with pure photoionized gas sitting on the second track at larger temperatures \citep[e.g.,][]{Rey:2025}. Continuing to the strongest leakers ($f_{\rm esc}\geq50\%$), we find a completely different ISM structure. In all cases, we find that photoionization heating (especially in dense environments) has been incredibly impactful, heating a significant mass fraction of the ISM to close to $\sim 30,000~{\rm K}$. Among these galaxies, the cold ($T\sim 100~{\rm K}$), dense, star-forming ISM is almost effectively depleted -- a strong signature of SNe feedback that has effectively coupled to the ISM, heating gas to $\sim 10^6~{\rm K}$ \citep{Gent:2012,Kim:2017}. Furthermore, we note that the \SF\ and \ISM\ simulations are capable of efficient ejective feedback, clearing gas beyond the central 25\% of the virial radius (see Figure 8 of \citealt{Attard:2026}). Interestingly, the dense photoionized line-emitting gas ($n_{\rm H}\gtrsim 10^3~{\rm cm^{-3}}$ and $T\sim 10^4~{\rm K}$, see Figure 2 of \citealt{Choustikov:2025b}) has also been depleted, consistent with the fact that extremely strong LyC leakers show weak emission line equivalent widths \citep[e.g.,][]{Nakajima:2014,Steidel:2018,Choustikov2024,Choustikov:2024b}.

\begin{figure}
    \centering
    \includegraphics[]{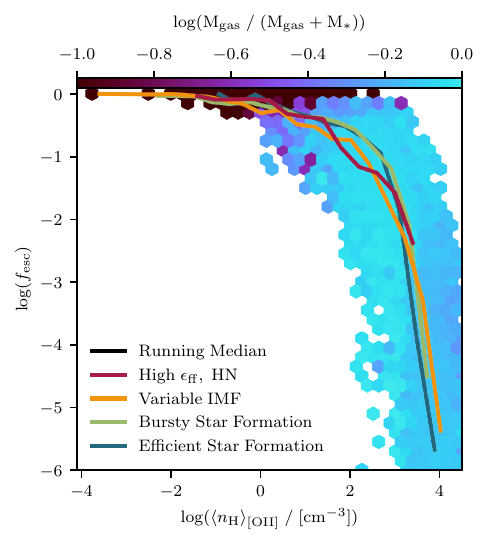}
    \caption{\textbf{Connecting LyC escape to gas density.}\\
    We show $f_{\rm esc}$ as a function of the $\rm [O~{\small II}]~\lambda\lambda3727$-weighted hydrogen gas density for galaxies across all {\small MEGATRON} simulations with $\rm M_{\rm UV}\leq -15$. Each pixel is coloured by the median gas fraction of individual galaxies, highlighting that halos with $f_{\rm esc}\sim100\%$ have expelled a significant portion ($\gtrsim90\%$) of their gas. Running medians are provided for each individual simulation, highlighting that while the distribution of escape fractions is strongly modified by feedback physics, the connection between $f_{\rm esc}$ and $\langle n_{\rm H}\rangle_{\rm [OII]}$ remains unchanged.}
    \label{fig:fesc_nH_O2}
\end{figure}

\begin{figure*}[!t]
    \centering
    \includegraphics[width=0.8\textwidth]{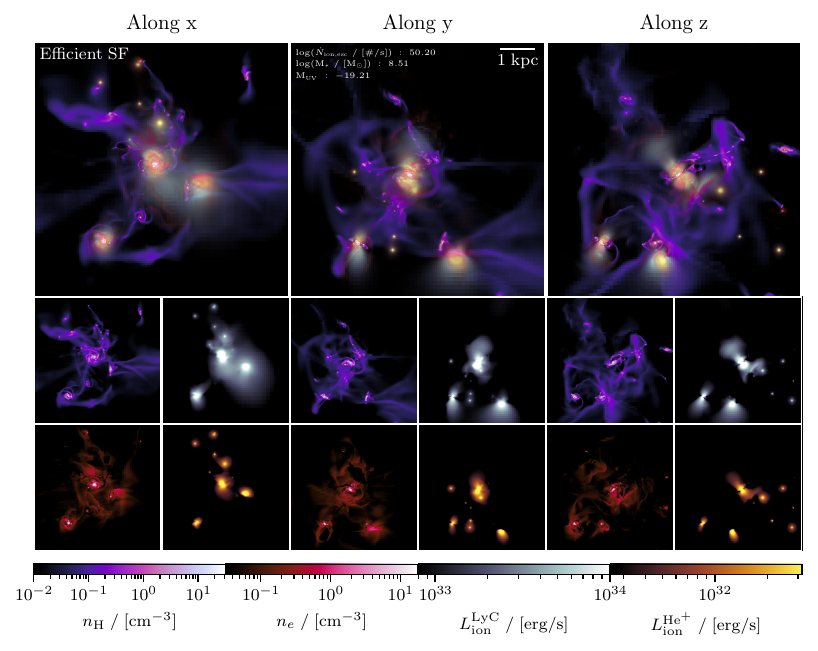}
    \includegraphics[width=0.8\textwidth]{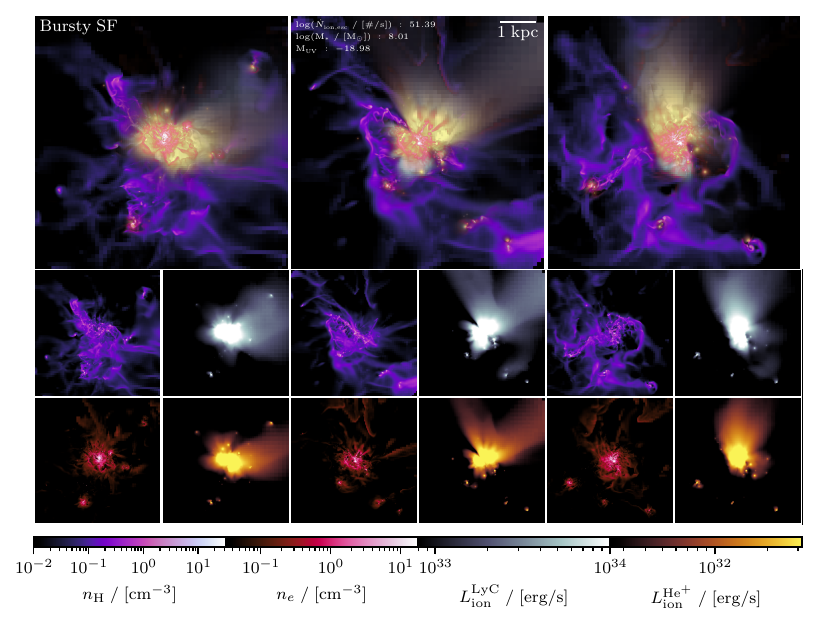}
    \caption{\textbf{Anisotropic LyC emission from galaxies in the \textit{efficient/bursty star formation} simulations at redshift $\bm{z\sim 8.5}$.}\\
    We show projections of hydrogen gas density (purple), electron density (red), LyC luminosity (white), and $\rm He^+$-ionizing luminosity (yellow) along each axis of the most massive galaxy at redshift $z\sim8.5$ in the \OG\ (top) and \SF\ (bottom) simulations. In each case, a composite of the four quantities is also provided. In both simulations, we find that LyC leakage is highly anisotropic and frequently attenuated along sightlines with high gas column densities, where high electron number densities can be found.}
    \label{fig:nH_LyC_projections}
\end{figure*}

Building on our findings for the \OG\ simulation, we can repeat this exercise for the other {\small MEGATRON} simulations, shown in the other rows. Despite the fact that the intrinsic ISM structure and $f_{\rm esc}$ distributions (see Figure \ref{fig:fesc_xi_nion_hist}) are completely different across the suite, we find that the trends described above are consistent. In general, strong photoionization feedback is a hallmark of intermediate leakers ($1\%\leq f_{\rm esc}\leq10\%$), strong leakers ($10\%\leq f_{\rm esc}\leq50\%$) show a combination of effective photoionization and SNe feedback, and the strongest leakers ($f_{\rm esc}\geq50\%$) show a lack of dense star-forming and line-emitting gas, with effectively all of the gas having been photoionized and cleared by SNe. This is a crucial finding, as it shows that while changes to feedback physics can modify $f_{\rm esc}$, the mechanisms by which it achieves the production of escape channels through the dense ISM is unchanged.

\begin{figure}
    \centering
    \includegraphics[]{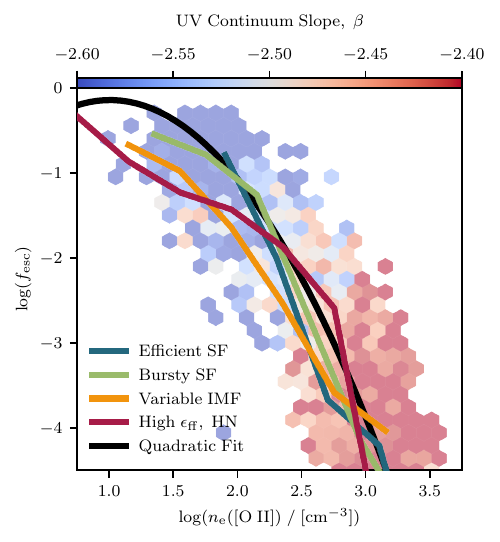}
    \caption{\textbf{Predicting LyC escape with electron density.}\\
    We show $f_{\rm esc}$ as a function of electron number densities measured from the $\rm [O~{\small II}]~\lambda\lambda3727$ emission line doublet, with individual pixels coloured by the UV continuum slope, $\beta$. We include all galaxies with $\rm M_{\rm UV}\leq -17$ and $\rm EW(H\beta)>50\AA$ in order to produce an observationally comparable sample. Running medians are provided for each individual simulation, alongside a quadratic fit as given in Equation \ref{eq:fesc_neO2_fit}. We find this trend continues to be robust to stellar feedback physics, highlighting that $n_e(\rm [O~{\small II}])$ contains useful information about the leakage of ionizing radiation.}
    \label{fig:fesc_ne_O2}
\end{figure}

\begin{figure*}
    \centering
    \includegraphics[width=\textwidth]{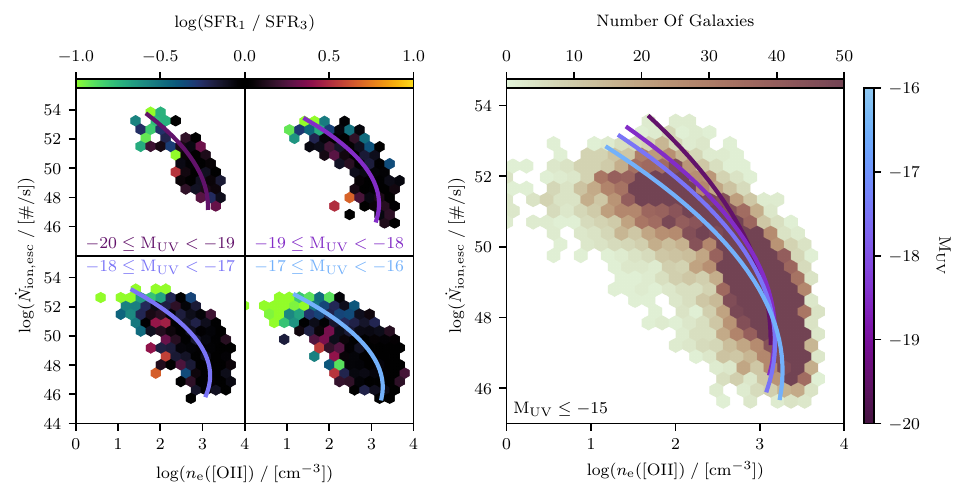}
    \caption{\textbf{Connecting the escaping ionizing luminosity to emission line-derived ISM electron densities.}\\
    We show $\dot{N}_{\rm ion,esc}$ as a function of electron densities measured from the $\rm [O~{\small II}]~\lambda\lambda3727$ emission line doublet and coloured by the burstiness of the recent SFH (defined using $\rm SFR_1/SFR_3$). On the left, each panel shows a subset of galaxies in a specific UV magnitude range. We only include galaxies with $\rm EW(H\beta)\ge 50\AA$ to account for the difficulty of measuring ISM properties from weak emission lines. We fit quadratics of the form $\log(n_e({\rm [O~{\small II}]}\;/\;[{\rm cm^{-3}}])) = c_0 + c_1y + c_2y^2$ (where $y = \log(\dot{N}_{\rm ion,esc}\;/\;[\#/{\rm s}])$) to each subset of galaxies, with coefficients given in Table \ref{tab:Nion_ne_O2_fit}. Finally, we show the distribution of all galaxies with $\rm M_{\rm UV}\leq-15$ and $\rm EW(H\beta)\ge 50\AA$ in this space, highlighting each of the fits. In general, galaxies with low electron densities ($\lesssim 300\,{\rm cm^{-3}}$) and recently declining SFHs tend to release more ionizing photons into the IGM, with UV-bright galaxies showing less scatter in the trend. Nevertheless, these functions are not strongly predictive, meaning that they should mainly be used for population-level analyses.}
    \label{fig:Nion_ne_O2}
\end{figure*}

\begin{table}[!h]
    \centering
    \caption{Fitted functions for lines shown in Figure \ref{fig:Nion_ne_O2}. All functions are of the form $x = c_0 + c_1y + c_2y^2$, where $x = \log(n_e({\rm [O~{\small II}]})\;/\;[{\rm cm^{-3}}])$ and $y = \log(\dot{N}_{\rm ion,esc}\;/\;[\#/{\rm s}])$.}
    \begin{tabular}{cccc}
         $\rm M_{\rm UV}$ Range & $c_0$ & $c_1$ & $c_2$ \\
         \hline
         $-20\leq \rm M_{\rm UV}<-19$ & $-77.52$ & $3.406$ & $-0.03595$  \\
         $-19\leq \rm M_{\rm UV}<-18$ & $-112.9$ & $4.885$ & $-0.05137$  \\
         $-18\leq \rm M_{\rm UV}<-17$ & $-116.6$ & $5.068$ & $-0.05359$  \\
         $-17\leq \rm M_{\rm UV}<-16$ & $-108.9$ & $4.853$ & $-0.05190$  \\
         \hline
    \end{tabular}
    \vspace{0.8cm}
    \label{tab:Nion_ne_O2_fit}
\end{table}

To further highlight this, Figure \ref{fig:integrated_phase_diagrams} shows an integrated variant of the same phase diagrams. Specifically, the left/right columns show density/temperature distributions from the weakest leakers (red) to the strongest (blue). Again, we find that while the structure of the ISM is different between the four {\small MEGATRON} simulations, the signatures of higher $f_{\rm esc}$ are the same: less cold neutral/dense gas ($T\lesssim10^3~{\rm K}$ and $n_{\rm H}\gtrsim10^2~{\rm cm^{-3}}$) and more warm ionized/low-density gas ($T\gtrsim 10^{4.5}~{\rm K}$ and $n_{\rm H}\lesssim 1~{\rm cm^{-3}}$). Moreover, it is interesting to note that in contrast to the other simulations, some dense gas remains intact even in extreme leakers ($f_{\rm esc}\geq50\%$) in the \OG\ simulation. Looking towards the future, it would also be interesting to explore the turbulent properties of each ISM phase in LyC-leaking galaxies, as under idealised conditions the presence of turbulence can occasionally produce low-density channels, promoting leakage without the need for any feedback whatsoever (e.g., \citealt{Xing:2026}, see also discussion by \citealt{Kakiichi2021}). Furthermore, early feedback is more effective at ejecting and ionizing lower-density channels in a turbulent medium while leaving higher-density channels unaffected \citep[e.g.,][]{Thompson:2016,Menon:2025}, thereby increasing the patchiness of the ISM and associated LyC escape. We however leave such discussions to future work, as {\small MEGATRON} is unable to resolve such physics at present.

To consolidate these findings, Figure \ref{fig:fesc_nH_O2} shows $f_{\rm esc}$ as a function of the hydrogen gas density weighted by $\rm [O~{\small II}]~\lambda\lambda3727$, $\langle n_{\rm H}\rangle_{\rm [OII]}$. Each pixel shows the median gas fraction for galaxies across all four {\small MEGATRON} simulations with $\rm M_{\rm UV}\leq-15$. Crucially, we find that all four simulations reproduce the same running median -- confirming that gas density contains some information about the leakage of ionizing radiation. Interestingly, we find that hydrogen density weighted by either the $\rm [O~{\small II}]~\lambda\lambda3727$ or $\rm [S~{\small II}]~\lambda\lambda6717,6731$ doublets carry considerably more information about $f_{\rm esc}$ than densities weighted by $\rm C~{\small III}]~\lambda\lambda1907,1909$, or $\rm N~{\small IV}]~\lambda\lambda1483,1486$. The reason for this is likely due to the fact that $\rm [O~{\small II}]$ and $\rm [S~{\small II}]$ are preferentially sensitive to lower-density gas (see discussions in \citealt{Kewley:2019b,Choustikov:2025b}). Furthermore, for a given density profile, a low electron density measured from $\rm [O~{\small II}]$ implies a larger ionized region due to the fact that $\rm O^+$ is a boundary species owing to the charge exchange interaction with hydrogen \citep{Jin2022,Xing:2026}. Altogether, $\langle n_{\rm H}\rangle_{\rm [OII]}$ carries more information about the density structure on the edge of the $\rm H~{\small II}$ regions. Finally, we find that the pure leakers with $f_{\rm esc}\sim100\%$ tend to be extremely gas-poor, having almost completely evacuated their ISM within 25\% of the virial radius (see also Figure 8 of \citealt{Attard:2026}). Such pure leakers tend to have weak emission line equivalent widths as well as high values of $\rm [O~{\small III]~\lambda5007}/[O~{\small II}]~\lambda\lambda3727$ for a given gas density (see Figure 9 of \citealt{Choustikov:2025b}). 

A visual representation of this discussion is presented in Figure \ref{fig:nH_LyC_projections}. Here, we show projections along each axis in gas density (purple), electron density (red), LyC luminosity (white) and $\rm He^+$-ionizing luminosity (yellow) for the most massive galaxy at redshift $z\sim8.5$ in the \OG\ (top) and \SF\ (bottom) simulations. In the composite images, we find that LyC leakage is highly anisotropic in both simulations, in agreement with previous studies \citep{Gnedin:2008,Wise2009,Paardekooper:2015,Trebitsch:2017,Kim:2023,Choustikov:2024b,Solhaug:2025,Ejdetjarn:2026}. Sightlines featuring high hydrogen column densities readily absorb ionizing flux, leaving behind regions of high electron number densities. Furthermore, it is important to note that there are significant morphological differences in the galaxies produced by these models (see discussion in Matsumoto et al. \textcolor{blue}{\textit{in prep.}}). For example, galaxies in the \OG\ simulation feature cold, dense discs with high electron number densities and biconical LyC escape. In contrast, galaxies in the \SF\ simulation are considerably more clumpy and disturbed by strong feedback, with LyC absorption occurring also due to filaments and ejected material in the CGM. This trend agrees with previous work with the {\small SPICE} simulations, which showed that dispersion-dominated galaxies tend to feature systematically larger escape fractions \citep{Bhagwat:2024b}. Finally, it is interesting to note that some galaxies in the \OG\ simulation leak $\rm He^+$-ionizing radiation in only a subset of LyC-leaking sightlines. As the former is hard radiation (here\footnote{In reality, there are a number of other mechanisms to produce $\rm He^{+}$-ionizing photons that are not included in {\small MEGATRON}. These include X-ray binaries \citep[e.g.,][]{Schaerer:2019}, very massive stars \citep[e.g.,][]{Schaerer:2025}, and AGN \citep[e.g.,][]{Feltre:2016}.} produced by lower-metallicity young massive stars, \citealt{Xiao2018,Stanway:2019,Saxena:2020}), this implies that such galaxies are releasing LyC late into a starburst cycle, as demonstrated in Figures \ref{fig:time_evo} and \ref{fig:sfr_stacks}.

\begin{figure*}
   \centering
   \includegraphics[width=\textwidth]{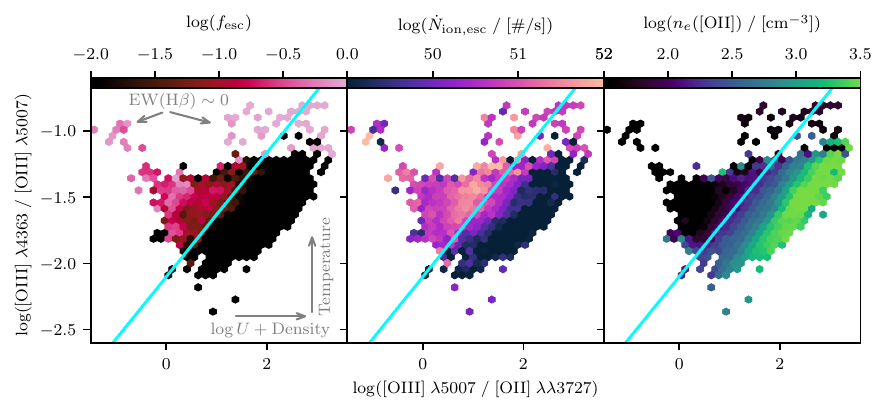}
    \caption{\textbf{Identifying LyC leakers with ISM properties: the TUne diagram.}\\
    We provide a line ratio diagnostic diagram of all galaxies with $\rm M_{\rm UV}\leq -15$ across each {\small MEGATRON} simulation. Specifically, we use the temperature-sensitive $\rm O33 = [O~{\small III}]~\lambda4363/[O~{\small III}]~\lambda5007$ and ionization parameter/density-sensitive $\rm O32 = [O~{\small III}]~\lambda5007/[O~{\small II}]~\lambda\lambda3727$ ratios, colouring each pixel by the median value of $f_{\rm esc}$ (left) and $\dot{N}_{\rm ion,esc}$ (centre), and $n_e([{\rm OII}])$ (right). Significant outliers have low gas fractions, leading to extremely weak emission line equivalent widths. Finally, we indicate a potential helpful but insufficient classifier in cyan given by $\rm \log(O33)>0.47\log(O32) - 2.1$. This selection produces a sample of which $>50\%$ of galaxies have $f_{\rm esc}>10\%$, capturing $88\%$ of all such halos. Similarly, this selection captures the majority of galaxies with $\dot{N}_{\rm ion,esc}\gtrsim10^{52}~\rm s^{-1}$, though galaxies with slightly larger values of O32 are still able to contribute to reionization, despite featuring LyC escape fractions below $\sim 1\%$.}
    \label{fig:O32_O33}
   \includegraphics[width=0.9\textwidth]{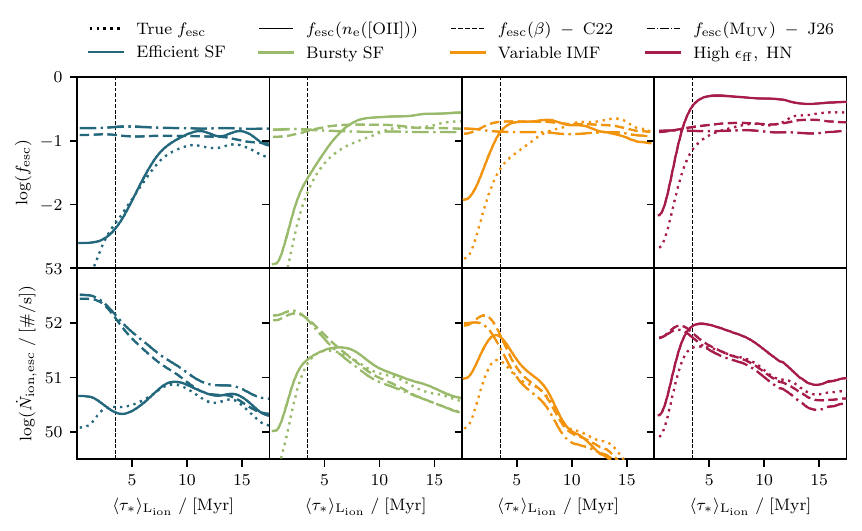}
   \caption{\textbf{Time-evolution of LyC leakage from galaxies using a variety of indirect models for the escape fraction.}\\
   We show running means (smoothed with a Gaussian with standard deviation $1~\rm Myr$) of $f_{\rm esc}$ and $\dot{N}_{\rm ion,esc}$ as a function of the ionizing luminosity-weighted mean stellar age for galaxies in each {\small MEGATRON} simulation with $\rm M_{\rm UV}\leq-13$. We include lines based on the true escape fraction (dotted), as well as based on indirect methods such as the O2-inferred electron number density (solid, Equation \ref{eq:fesc_neO2_fit}), the UV continuum slope (dashed, \citealt{Chisholm2022}), and the UV magnitude (dot-dashed, \citealt{Jecmen:2026}). We find that $f_{\rm esc}(\beta)$ and $f_{\rm esc}(\rm M_{\rm UV})$ are constant with time, compared to the significant early time evolution in the true $f_{\rm esc}$. This leads to systematic over-predictions in $\dot{N}_{\rm ion,esc}$ by up to two orders of magnitude. In contrast, $f_{\rm esc}(n_e([\rm O~{\small II})])$ is capable of reproducing much of this early time evolution, especially when applied to the \OG\ and \SF\ simulations. While the ISM density-based approach systematically over-predicts $\dot{N}_{\rm ion,esc}$ at late times in the \ISM\ simulation, this is a result of the fact that this model produces galaxies with a characteristically less dense ISM, therefore biasing the inferred escape fractions.}
   \label{fig:beta_time_evo}
\end{figure*}

\section{Probing Galactic Contributions to Reionization with ISM Density}
\label{sec:LyC_diagnostics}

\noindent Given the findings of Section \ref{sec:ISM_structure}, we now proceed to investigate whether directly observable probes of ISM properties might carry information about $f_{\rm esc}$ and $\dot{N}_{\rm ion,esc}$.

\subsection{ISM Density as a Probe of LyC Escape}

\noindent To begin, Figure \ref{fig:fesc_ne_O2} shows $f_{\rm esc}$ as a function of the electron number density inferred from the $\rm [O~{\small II}]~\lambda\lambda3727$ doublet, $n_e([{\rm O~{\small II}}])$, for galaxies with $\rm M_{\rm UV}\leq-17$ and $\rm EW(H\beta)>50\AA$ across each {\small MEGATRON} simulation. An added cut of $\rm EW(H\beta)\geq50\AA$ is included to account for the fact that it is difficult to reliably measure ISM properties from weak emission lines. We find that a trend does emerge, driven by the presence of strong leakers with $n_e([{\rm O~{\small II}}])\sim 30-100~{\rm cm^{-3}}$ and dust-free blue UV continuum slopes ($\beta\lesssim-2.6$, \citealt{Chisholm2022}) alongside non-leakers with $n_e([{\rm O~{\small II}}])\sim 1000~{\rm cm^{-3}}$ and red UV continuum slopes ($\beta\gtrsim-2.4$) driven by the nebular continuum \citep{Katz2024_BJ}. Interestingly, we find several galaxies with blue UV slopes but low escape fractions. Such systems all feature ISM densities lower than the median relation and feature blue slopes due to a nebular continuum that has been weakened by non-equilibrium physics \citep{Katz:2026c}. 

Similar to the results of Figure \ref{fig:fesc_nH_O2}, we find that running medians from the four simulations tend to agree (albeit with comparatively more scatter), suggesting that there is still a consistent trend. A line of best fit is computed (amalgamating data from all four simulations) to capture this trend, given by:
\begin{equation}
    \label{eq:fesc_neO2_fit}
    \log(f_{\rm esc}) = -1.094 + 1.890x - 0.939x^2,
\end{equation}
where $x= \log(n_e([{\rm O~{\small II}}])\;/\;[{\rm cm^{-3}}])$. We re-iterate that this trend is not predictive, especially when compared to published multi-variate relations for $f_{\rm esc}$ \citep[e.g.,][]{Mascia:2023,Choustikov2024,Choustikov:2024b,Jaskot:2024,Jaskot:2024b} and $\dot{N}_{\rm ion,esc}$ \citep{Choustkov:2025,Summerfield:2026}. Furthermore, there is reason to expect that this relation should weaken in more massive systems. Here, the combination of dust and the presence of multiple emitting regions are likely to wash out this tracer to a typical galactic-averaged value of $n_e([{\rm O~{\small II}}])\sim 300-1000~{\rm cm^{-3}}$ as line emission is dominated by dense gas (\citealt{Xing:2026}, see also discussion in \citealt{Katz:2026c}). Similarly, we find that while $n_e({\rm [O~{\small II}]})$ traces the global (angle-averaged) escape fraction, the trend does not hold for line-of-sight escape fractions (see discussion in Appendix \ref{app:ne_fesc_SPHINX}). Nevertheless, $n_e([{\rm O~{\small II}}])$ is potentially a useful indicator when applied to population-scale samples, especially given published trends for the observed redshift evolution of ISM densities \citep[e.g.,][]{Isobe2023,Abdurro'uf:2024,Topping2025,Martinez:2025,Stephenson:2026}.

We have also investigated a number of density-sensitive doublets as potential tracers of $f_{\rm esc}$. We find that indicators sensitive to low-density gas (e.g., $\rm [S~{\small II}]~6716/6731$, $\rm [O~{\small III}]~88\mu m/52\mu m$, and $\rm [S~{\small III}]~33.5\mu m/18.7\mu m$) can also provide insight about LyC escape (see discussion in Xing et al., \textcolor{blue}{\textit{in prep.}}). However, these ratios (especially $\rm [S~{\small II}]$) are less robust to feedback physics due to contributions from the diffuse ionised gas. In contrast, UV tracers of dense gas (e.g., $\rm C~{\small III}]~1906/1908$ and $\rm N~{\small IV}]~1483/1486$) are not as sensitive to the effective coupling of feedback in the ISM, as they are biased to extremely dense environments \citep{Topping2025,Martinez:2025,Berg:2026}.

Spurred on by these findings, we investigate the dependence of $\dot{N}_{\rm ion,esc}$ on $n_e([{\rm O~{\small II}}])$ in Figure \ref{fig:Nion_ne_O2}. On the left, each panel shows this relation for a different bin in absolute UV magnitude, coloured by the burstiness of the recent SFH (defined using $\rm SFR_1/SFR_3$). Furthermore, we retain the cut of $\rm EW(H\beta)\geq50\AA$ as discussed above. We find that in all cases, there is a trend wherein galaxies which appear to have a low-density ISM and a recently declining SFH are likely to be releasing the most ionizing photons into the IGM (in agreement with our findings in Sections \ref{sec:reionization} and \ref{sec:ISM_structure}. Interestingly, galaxies with $-20\leq\rm M_{\rm UV}\leq-19$ show the tightest relation. In contrast, fainter galaxies feature a more shallow slope with increased scatter. For each UV magnitude bin, we fit a power law of the form $x = c_0 + c_1y + c_2y^2$, where $y = \log(\dot{N}_{\rm ion,esc}\;/\;[\#/{\rm s}])$ and $x = \log(n_e([{\rm O~{\small II}}])\;/\;[{\rm cm^{-3}}])$ using the same approach that is typically applied to strong-line metallicity calibrations (e.g., \citealt{Nakajima:2022b,Laseter2024,Sanders:2024,Sanders:2026,Isobe:2026}, Choustikov et al., \textcolor{blue}{\textit{in prep}.}). Each line is displayed in the relevant panel of Figure \ref{fig:Nion_ne_O2} and compiled on the rightmost panel alongside the underlying distribution of all galaxies with $\rm M_{\rm UV}\leq-15$ and $\rm EW(H\beta)\geq50\AA$. Coefficients are provided in Table \ref{tab:Nion_ne_O2_fit}, though it is important to highlight that the fits provided in Table \ref{tab:Nion_ne_O2_fit} are not strongly predictive (especially at the high-density limit). This means that they should be applied with caution and are best-used for population-level analyses.

\subsection{Ambient ISM Properties of LyC-Leaking Galaxies: The TUne Diagram}

\noindent Building on our findings that electron densities contain information about $f_{\rm esc}$ and $\dot{N}_{\rm ion,esc}$, we proceed to explore a multi-variate approach (see also \citealt{Choustikov2024}. To this end, we introduce the temperature-ionization parameter-density (TUne) diagram. Figure \ref{fig:O32_O33} shows $\rm O33 = [O~{\small III]~\lambda 4363/[O~{\small III}~\lambda 5007]}$ (a tracer of ISM temperature, \citealt{Kewley2019}) as a function of $\rm O32 = [O~{\small III]~\lambda 5007/[O~{\small II}~\lambda\lambda 3727]}$ (a tracer of ionization parameter and density, \citealt{Martinez:2025}) for galaxies with $\rm M_{\rm UV}\leq-15$ across all four {\small MEGATRON} simulations, coloured by the median $f_{\rm esc}$ (left),  $\dot{N}_{\rm ion,esc}$ (centre), and $n_e({[\rm OII]})$ (right) of each pixel. We find that galaxies with $f_{\rm esc}\gtrsim10\%$ and $\dot{N}_{\rm ion,esc}\gtrsim10^{50.5}~\rm s^{-1}$ are confined to a particular portion of the diagram corresponding to a hot, low density ISM (consistent with the findings of Figures \ref{fig:full_phase_diagrams} and \ref{fig:integrated_phase_diagrams}). Furthermore, the strongest leakers ($f_{\rm esc}\sim100\%$) have an extremely hot ISM, though much of this is driven by them also being very gas poor (making emission lines unobservable). An example visually defined boundary is shown in cyan, given by:
\begin{equation}
    \log(\rm O33) > 0.47\times \log(O32) -2.1.
\end{equation}
This helpful but insufficient criterion produces a sample of which 52\% (14\%) have $f_{\rm esc}>10\%$ ($\dot{N}_{\rm ion,esc}>10^{52}~\rm s^{-1}$) and that manages to capture 90\% (82\%) of all such systems across all four {\small MEGATRON} simulations. Looking to the central panel, we find that galaxies releasing the most ionising photons into the IGM tend to have $\rm O32\sim 5-15$ (consistent with the findings of \citealt{Choustikov2024}). At lower O32, $\dot{N}_{\rm ion,esc}$ falls as fewer ionizing photons are being produced, while at larger O32 the dense ISM has not yet been cleared and $\rm [O~{\small II}]~\lambda\lambda3727$ is collisionally de-excited. Nevertheless, we find that for galaxies with a fixed value of O32 ratios within this range, the temperature of the ISM is a strong indicator of whether a galaxy is effectively leaking LyC photons.

\subsection{Time-Evolution of Observational Diagnostics of LyC Escape}

\noindent Another interesting potential benefit of ISM structure-based tracers is their potential ability to track the time-evolution of $f_{\rm esc}$ over an individual burst cycle. For example, the electron density of a starburst will track both the ionization parameter as well as the process of feedback. This combination should therefore be sensitive to both density-bound and ionisation-bound leakage modes \citep{Zackrisson2013}. Similarly, O33 is indicative of the high-energy ionisation parameter (e.g., traced by $\rm C~{\small IV}~\lambda\lambda1550/C~{\small III}]~\lambda\lambda1908$, \citealt{Berg:2021}), as shown by Figure 9 of Choustikov et al., (\textcolor{blue}{\textit{in prep}.}).

To investigate this aspect, Figure \ref{fig:beta_time_evo} shows smoothed medians for $f_{\rm esc}$ and $\dot{N}_{\rm ion,esc}$ (using the true intrinsic $\dot{N}_{\rm ion,int}$) as a function of ionising luminosity weighted stellar age. We include smoothed (over a Gaussian of standard deviation $1~{\rm Myr}$) running medians considering the true LyC escape fraction (dotted), the $n_e([\rm O~{\small II}])$-based escape fraction given by Equation \ref{eq:fesc_neO2_fit} (bold), the $\beta$-based values using Equation 11 of \citet{Chisholm2022}, and the recent UV magnitude-based calibration\footnote{It is worth pointing out that this approach is built on a dataset of galaxies for which escape fractions are computed using the UV continuum slope \citep{Chisholm2022}.} using galaxies from the GLIMPSE survey \citep{Jecmen:2026}. Here, we find that $f_{\rm esc}(\beta)$ and $f_{\rm esc}(\rm M_{\rm UV})$ are both roughly constant in time, with the former reproducing the population average value of $\langle f_{\rm esc}\rangle = 0.14$ reported by \cite{Jecmen:2026}. In contrast, we find that $f_{\rm esc}(n_e([\rm O~{\small II}]))$ begins to capture the time-dependent nature of $f_{\rm esc}$ in {\small MEGATRON} galaxies. While this particular relation is less accurate for the \IMF\ and \ISM\ simulations (owing to their inclusion of hypernovae), the overall mean time evolution of $\dot{N}_{\rm ion,esc}$ is much better captured than by the traditional approach of \cite{Chisholm2022}. Finally, we note that while this method would over-predict the escaping ionising luminosity for the \ISM\ simulation, this is not unexpected given the fact that the ISM of this model is characteristically less dense (see Figure 7 of \citealt{Choustikov:2025b}). 

Altogether, we emphasise that the early time-evolution of $f_{\rm esc}$ and $\xi_{\rm ion}$ of UV-faint galaxies must be jointly accounted for in models of the reionization process. Another potential avenue to accomplish this may be to condition traditional escape fraction diagnostics on the inferred recent star formation history. We leave such discussions to future work.

\section{Conclusions}
\label{sec:conclusions}

\noindent Understanding the process of cosmic reionization -- and by extension -- the escape of Lyman Continuum (LyC) photons from star-forming galaxies is one of the key questions of modern astrophysics. In particular, it is vital to reconcile observed high-redshift UV luminosity functions and ionizing production efficiencies of early stellar populations with reionization histories that have been measured from parallel probes. Here, LyC escape fractions continue to be the main degree of uncertainty -- traditionally assumed constant or inferred by indirect diagnostics that have been calibrated on low-redshift analogues. 

To address this challenge, we conduct a study of the process of reionization in the {\small MEGATRON} simulations \citep{Katz:2025b,Cadiou:2025,Rey:2025b}: a suite of high-resolution cosmological zoom-in simulations of a Milky Way-mass progenitor that feature on-the-fly radiation transfer coupled to an extensive non-equilibrium thermochemical network (81 species) and detailed model for heating and cooling processes in the ISM. In particular, we utilise a number of re-simulations which feature modifications to star-formation/feedback physics, inspired by various potential solutions to the observed number of UV-bright galaxies and capable of reproducing the observed ISM properties of high-redshift galaxies \citep{Choustikov:2025b}. With this as a basis, we investigate both the demographics and ISM structure of LyC-leaking halos. Our results are as follows:

\begin{itemize}
   \item \textbf{Number counts of UV sources are directly linked to the distribution of LyC escape fractions by feedback physics.} Changes to star formation/feedback physics will directly mediate the number of galaxies at a given $\rm M_{\rm UV}$, the contribution of each galaxy to the ionizing photon background, and ultimately the reionization history as a whole. As a result, the presence of a large number of UV sources does not imply a commensurate increase in the total escaping ionizing luminosity, with the two features appearing to anti-correlate.
   
   \item \textbf{Understanding the time-delay between a starburst and the onset of effective LyC escape is the key challenge.} The primary physical factor connecting the number of galaxies to the ionizing photon budget is the relative timescale of processes that allow LyC photons to escape compared to the lifetime of LyC-producing stellar populations. More effective early feedback (e.g., radiation and hypernovae) released in lower-density environments leads to a scenario where a smaller number of galaxies are capable of driving an earlier reionization. In contrast, efficient star formation in dense environments leads to longer delays between a starburst and effective leakage -- leading to a slow reionization irrespective of an increase to UV source number counts.

   \item \textbf{UV-faint LyC leakers tend to feature post-starburst star formation histories.} Similarly, UV-faint galaxies with rapidly rising, starburst, SFHs do not leak LyC photons. When the production efficiency of ionizing radiation is accounted for, post-starburst systems are still the dominant UV-faint LyC leakers.
   
   \item \textbf{The characteristic ISM structure of UV-faint LyC leakers is similar across models with different feedback physics.} While changes to feedback fundamentally modify the structure of the ISM, the physical mechanisms that drive LyC escape are similar in all four simulations. Strong photoionization feedback is a hallmark of intermediate ($1\%\leq f_{\rm esc}\leq 10\%$) leakers, strong leakers ($10\%\leq f_{\rm esc}\leq 50\%$) show a combination of effective coupling from photoionization and SNe feedback, while extreme leakers ($f_{\rm esc}\geq50\%$) show a lack of dense star-forming and line-emitting gas. This leads to a relationship between $f_{\rm esc}$ and the average hydrogen gas density that is consistent across all four simulations.
   
   \item \textbf{ISM Electron number density measured from the $\rm [O~{\small II]~\lambda\lambda3727}$ emission line doublet is a potential indicator of LyC escape in UV-faint galaxies.}
   Based on our finding that gas density contains information about the production of escape channels, we propose that the electron number density measured from the $\rm [O~{\small II}]~\lambda\lambda3727$ doublet may be a useful indicator of $f_{\rm esc}$ and $\dot{N}_{\rm ion,esc}$ in UV-faint galaxies. Crucially, this trend appears to be robust to changes to feedback physics. While these trends are not strictly predictive, they may provide a viable method to identify leakers among large populations of galaxies. Furthermore, we note that $n_e({[\rm O~{\small II}]})$ can be polluted by emission from multiple emitting regions (e.g., in more massive galaxies) and is generally more indicative of global $f_{\rm esc}$ as compared to the value measured along a single sight-line.

   \item \textbf{At a fixed ionization parameter, the ISM of LyC leakers is hotter and less dense.} This is summarised in the TUne diagram, which combines $\rm O33 = [O~{\small III}]~\lambda4363/[O~{\small III}]~\lambda5007$ and $\rm O32 = [O~{\small III}]~\lambda 5007/[O~{\small II}]~\lambda\lambda3727$. Galaxies with high escape fractions cluster in a region of the O33-O32 space that is associated with lower electron densities and high temperatures. Furthermore, galaxies with large escaping ionizing luminosities tend to show larger values of O32 but still fall in the same region.
   
   \item \textbf{Traditional UV continuum slope-based escape fraction calibrations over-predict escaping ionizing luminosities by failing to capture the time evolution of the feedback processes.} In contrast to traditional indicators based on the UV continuum slope ($\beta$), $n_e([\rm O~{\small II}])$ is able to better capture the time evolution of $f_{\rm esc}$ and $\dot{N}_{\rm ion,esc}$. At early times ($\tau_* \lesssim 5~{\rm Myr}$), the $f_{\rm esc}(\beta)$ calibration of \cite{Chisholm2022} may over-predict the luminosity of escaping ionising radiation from individual sources by $\sim 2$ orders of magnitude. However, this calibration tends to the correct value at later times. \textbf{The delayed duty cycle of LyC leakage must be accounted for in models of reionization that are based on observed UV source number counts.}
\end{itemize}

This work has highlighted that reionization-relevant observables such as UV luminosity functions, ionizing photon production efficiencies, and LyC escape fractions are inextricably linked by time-delayed duty cycles informed by feedback physics. Furthermore, we argue that tracers of the multi-phase ISM structure such as electron densities and temperatures should be explored as potential $f_{\rm esc}$ indicators -- especially in the era of exquisite spectra observed by \textit{JWST}. Nevertheless, future work will be needed to characterise the impact of other feedback channels (e.g., $\rm Ly\alpha$ pressure, cosmic rays, and AGN) on the time-dependent process of Lyman Continuum escape. Furthermore, we will ultimately need to address the divide between ultra-high resolution GMC-scale star formation simulations and cosmological simulations of reionization with a resolved multi-phase ISM. {\small MEGATRON} represents good progress in this direction, but we must continue to leverage the synergy between high-resolution simulations, synthetic observations, and state-of-the-art observational facilities to understand the contributions of individual galaxies to cosmic reionization.

\section*{Data Availability}

\noindent The data employed in this manuscript will be shared upon reasonable request to the corresponding author.

\section*{Acknowledgements}

\noindent The authors would like to thank Lily Whitler and Christopher Cain for insightful discussions.

NC acknowledges support from the Kavli Institute for Cosmology, Cambridge and the Kavli Foundation. TK were supported by the National Research Foundation of Korea (RS-2022-NR070872 and RS-2025-00516961). MJH is fellow of the Knut \& Alice Wallenberg Foundation. MPR acknowledges support from UKRI Science
and Technology Facilities Council (STFC) grant UKRI4103.
Support for program JWST-AR-08709 (AS) was provided by NASA through a grant from the Space Telescope Science Institute, which is operated by the Association of Universities for Research in Astronomy, Inc., under NASA contract NAS 5-03127.

Megatron made extensive use of the dp016, dp265, dp373, and dp379 projects on the DiRAC ecosystem. This work was performed using the DiRAC Data Intensive service at Leicester, operated by the University of Leicester IT Services, which forms part of the STFC DiRAC HPC Facility (www.dirac.ac.uk). The equipment was funded by BEIS capital funding via STFC capital grants ST/K000373/1 and ST/R002363/1 and STFC DiRAC Operations grant ST/R001014/1. This work used the DiRAC@Durham facility managed by the Institute for Computational Cosmology on behalf of the STFC DiRAC HPC Facility (www.dirac.ac.uk). The equipment was funded by BEIS capital funding via STFC capital grants ST/P002293/1, ST/R002371/1 and ST/S002502/1, Durham University and STFC operations grant ST/R000832/1. This work was performed using resources provided by the Cambridge Service for Data Driven Discovery (CSD3) operated by the University of Cambridge Research Computing Service (www.csd3.cam.ac.uk), provided by Dell EMC and Intel using Tier-2 funding from the Engineering and Physical Sciences Research Council (capital grant EP/T022159/1), and DiRAC funding from the Science and Technology Facilities Council (www.dirac.ac.uk). DiRAC is part of the National e-Infrastructure. The authors thank Jonathan Patterson for smoothly running the Glamdring Cluster hosted by the University of Oxford, where part of the data processing was performed. The authors also acknowledge financial support from Oriel College’s Research Fund. The material in this manuscript is based upon work supported by NASA under award No. 80NSSC25K7009.

The authors thank Romain Teyssier and Leo Michel-Dansac for both developing and open-sourcing {\small RAMSES} and {\small RASCAS}, respectively. We thank the developers and maintainers of \textsc{pynbody} (\citealt{Pontzen2013}), \textsc{yt} (\citealt{Turk2011}), \textsc{NumPy} (\citealt{vanderWalt2011, Harris2020}), \textsc{SciPy} (\citealt{Virtanen2020}), \textsc{jupyter} (\citealt{Ragan-Kelley2014}), \textsc{matplotlib} (\citealt{Hunter2007}), \textsc{cmasher} (\citealt{van-der-Velden:2020}), \textsc{pandas} (\citealt{reback2020pandas,mckinney-proc-scipy-2010}), the Astrophysics Data Service, and the arXiv pre-print repository for providing open-source software and services that were used extensively in this work.

\bibliographystyle{mn2e}
\bibliography{example}

\appendix

\section{A. Intrinsic ISM Gas Density and Observed Electron Density}
\label{app:nH_ne}

\begin{figure*}[t!]
   \centering
   \includegraphics[]{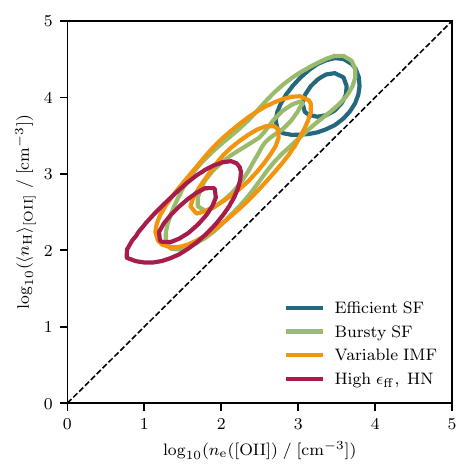}
   \caption{\textbf{A comparison between the intrinsic gas density and inferred electron density traced by $\bf{O^+}$.}\\
   We show $\langle n_{\rm H}\rangle_{\rm [O~{\small II}]}$ as a function of $n_e({\rm [O~{\small II}]})$ for all galaxies with $\rm M_{\rm UV}\leq-15$ across all four MEGATRON simulations.}
   \label{fig:app_neO2_nHO2}
\end{figure*}

In Figure \ref{fig:app_neO2_nHO2}, we show a comparison between the electron number density inferred from $\rm [O~{\small II}]~\lambda\lambda3727$ and the intrinsic average hydrogen gas density weighed by $\rm [O~{\small II}]$ for all galaxies with $\rm M_{\rm UV}\leq-15$ across each {\small MEGATRON} simulation. We find that there is a characteristic offset of $\sim 0.8~{\rm dex}$, with the electron density under-predicting the typical gas density of these regions. This is consistent with the fact that the $\rm [O~{\small II}]$ doublet is sensitive to a less-dense (owing to the low critical densities, \citealt{Kewley:2019b,Martinez:2025}) and lower ionization parameter regime.

\section{B. Impact of star-formation/feedback physics on scaling relations}
\label{app:scaling_relations}

\begin{figure*}[t!]
   \centering
   \includegraphics[width=\textwidth]{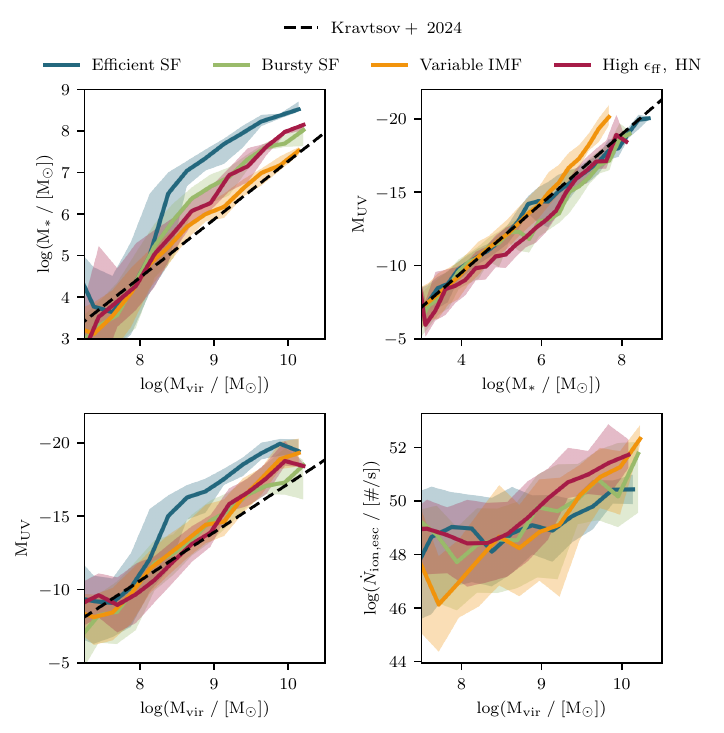}
   \caption{\textbf{Halo mass, stellar mass, UV magnitude, and escaping ionizing luminosity scaling relations in {\small MEGATRON}.}\\
   We show the joint scaling relations between the DM halo virial mass ($\rm M_{\rm vir}$), stellar mass (after mass loss, $\rm M_*$), UV magnitude ($\rm M_{\rm UV}$), and escaping ionizing luminosity $\dot{N}_{\rm ion,esc}$ for galaxies at redshift $8.5\leq z\leq 9.5$ across all four {\small MEGATRON} simulations. In each case, the running median is shown alongside the single standard deviation scatter. As a comparison point, we show the scaling relations produced by the semi-analytic model of \cite{Kravtsov:2024} that was able to reproduce observed UVLFs.}
   \label{fig:scaling_relations}
\end{figure*}

Median scaling relations for the halo mass ($\rm M_{\rm vir}$), stellar mass (after mass loss, $\rm M_*$), and UV magnitudes ($\rm M_{\rm UV}$) of galaxies at redshift $8.5\leq z\leq9.5$) across all four {\small MEGATRON} simulations are shown in Figure \ref{fig:scaling_relations}. We also include the median relation for $\dot{N}_{\rm ion,esc}$ as function of halo mass. As a comparison point, we show the scaling relations for a baryon conversion efficiency, $f_*=0.01$ of \cite{Kravtsov:2024}, a semi-analytic model that was able to reproduce observed UVLFs by leveraging scatter in the SFHs of galaxies (see also \citealt{Mason:2023}). The \SF\ and \ISM\ simulations produce galaxies of a similar stellar mass (and UV luminosity) for a given halo mass. In contrast, galaxies in the \OG\ simulation form stars much more efficiently (see discussion in \citealt{Katz:2025b}), while the considerable stellar mass loss due to feedback in the \IMF\ simulation leads to a lower normalisation. Similarly, the four models produce a very similar $\rm M_{\rm UV}-M_*$ relations (up to their scatter), with the \IMF\ simulation featuring a boost at the higher masses due to the assumed SPS models \citep{Leitherer1999,Eldridge2017,Stanway2018}. Finally, we find that at a fixed halo mass, there is a large degree of scatter in the escaping ionizing luminosity, with the four simulations' median relations agreeing more than expected.

\section{C. Robustness Test of Electron Density as a Tracer of Escape Fraction}
\label{app:ne_fesc_SPHINX}

\begin{figure*}[t!]
   \centering
   \includegraphics[]{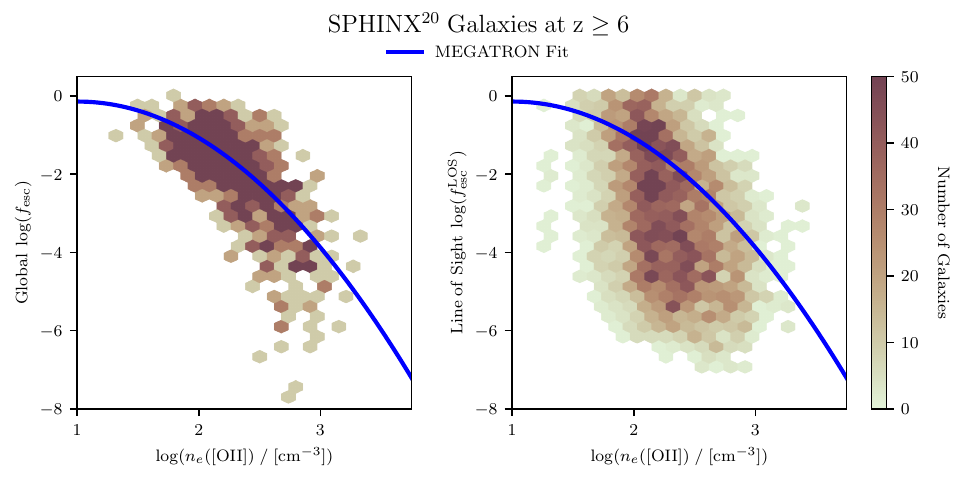}
   \caption{\textbf{Testing the robustness of the escape fraction -- density relation with the {\small SPHINX$^{\bm{20}}$} simulation.}\\
   Considering all galaxies with $\rm SFR_{10}\geq0.5~{\rm M_{\odot}yr^{-1}}$ at redshifts $z\geq6$ in the {\small SPHINX$^{20}$} simulation, we show the global LyC escape fraction (left) and line-of-sight LyC escape fraction (right) as a function of $n_e({[\rm O~{\small II}]})$. In each panel, we provide the best fit relation for {\small MEGATRON} as presented in Figure \ref{fig:fesc_ne_O2} and Equation \ref{eq:fesc_neO2_fit}. We find that this relation holds for the global escape fractions but disappears for line-of-sight escape fractions, owing to the highly anisotropic nature of LyC leakage.}
   \label{fig:app_fesc_ne_sphinx}
\end{figure*}

To further test the robustness of the relation between $f_{\rm esc}$ and $n_e([{\rm O~{\small II}}])$ proposed in Section \ref{sec:LyC_diagnostics}, we proceed to investigate this relation in galaxies from the {\small SPHINX$^{20}$} simulation \citep{Rosdahl2018,Rosdahl2022}. To do so, we use synthetic observations provided in the {\small SPHINX$^{20}$} Public Data Release v1 (SPDRv1, \citealt{Katz:2023}). We measure electron number densities from the intrinsic $\rm [O~{\small II}]~\lambda\lambda3727$ luminosities of the 696 galaxies at redshift $z\geq6$ with $\rm SFR_{10}\geq0.5~{\rm M_{\odot}yr^{-1}}$ using the method outlined in Section \ref{sec:methods}. Finally, we make use of both the global (i.e., angle-averaged) and line-of-sight LyC escape fractions ($f_{\rm esc}^{\rm LOS}$) that were computed using {\small RASCAS} (\citealt{Leo2020}, see discussions in \citealt{Choustikov2024,Choustikov:2024b}).

Figure \ref{fig:app_fesc_ne_sphinx} shows the global $f_{\rm esc}$ (left) and line-of-sight $f_{\rm esc}$ (right) as a function of $n_e([{\rm O~{\small II}}])$ for {\small SPHINX$^{20}$} galaxies. We find that  the trend with global $f_{\rm esc}$ holds. Given the fact that {\small SPHINX$^{20}$} uses a different stellar feedback prescription, a simpler treatment of heating and cooling physics in the ISM and slightly worse resolution of $\sim 10~\rm pc$ \citep{Rosdahl2018,Rosdahl2022}, this confirms that $f_{\rm esc}- n_e({[\rm O~{\small II}]})$ appears to be a robust trend in simulated galaxies during the Epoch of Reionization. Interestingly however, we find that no such trend exists when the line-of-sight escape fractions are considered. This speaks to the highly anisotropic nature of LyC escape (see Appendix B of \citealt{Choustikov:2024b} for a discussion of this in the context of {\small SPHINX$^{20}$}). Specifically, galaxies with global LyC escape fractions $>10\%$ are typically dominated by a few `successful' sightlines while also frequently featuring sightlines with $f_{\rm esc}^{\rm LOS}\sim0$ -- thereby weakening the relation. Another factor which should be considered here is the impact of morphology and ISM geometry (see discussions by \citealt{Bhagwat:2024b}), which would fundamentally alter the distribution of $f_{\rm esc}^{\rm LOS}$ for a given feedback event and ISM structure.

\end{document}